\documentclass[pdflatex,sn-mathphys-num,iicol]{sn-jnl}

\usepackage{graphicx}
\usepackage{amsmath,amssymb,amsfonts}
\usepackage{amsthm}
\usepackage{booktabs}
\usepackage{array}
\usepackage{xcolor}
\usepackage{colortbl}
\usepackage{etoolbox}
\definecolor{bestcell}{gray}{0.90}
\AtBeginEnvironment{table*}{\let\plainmathbf\mathbf\renewcommand{\mathbf}[1]{\cellcolor{bestcell}\plainmathbf{#1}}}
\usepackage{url}
\let\orcidlogo\undefined     
\usepackage{orcidlink}        
\usepackage{placeins}
\usepackage{float}
\usepackage{capt-of}            
\usepackage{rotating}           
\graphicspath{{figures/}}

\theoremstyle{thmstyleone}
\newtheorem{proposition}{Proposition}
\theoremstyle{thmstylethree}

\newcommand{\lgE}{\log_{10}(E/\mathrm{eV})}
\newcommand{\Ne}{N_{\mathrm{e}}}
\newcommand{\E}{\mathbb{E}}
\newcommand{\Rstar}{R^{\star}}

\newcommand{\Nroot}{171{,}776}

\newcommand{\sigmaCNNViT}{0.1063}

\newcommand{\kcdcDeltaMin}{-10}
\newcommand{\kcdcDeltaMax}{+1}

\newcommand{\kfSpread}{0.0013}

\newcommand{\nShowers}{47{,}394}
\newcommand{\meanReuse}{3.6}

\newcommand{\bayesFour}{0.1219}

\newcommand{\floorTest}{0.1018}
\newcommand{\floorTestLo}{0.1006}
\newcommand{\floorTestHi}{0.1030}

\newcommand{\cfGB}{0.1021}

\newcommand{\cfLeakMax}{0.0009}

\newcommand{\cfExcessAllMin}{0.2}
\newcommand{\cfExcessAllMax}{1.8}
\newcommand{\cfLadZen}{0.0042}
\newcommand{\cfLadMu}{0.0164}

\newcommand{\cfLadModLo}{-0.00003}
\newcommand{\cfLadModHi}{+0.00005}
\newcommand{\cfPermLo}{-0.00006}
\newcommand{\cfPermHi}{+0.00002}

\newcommand{\cfRendA}{0.1230}
\newcommand{\cfRendB}{0.1227}
\newcommand{\cfRendC}{0.1064}
\newcommand{\cfRendD}{0.1024}
\newcommand{\cfRendBA}{0.0003}
\newcommand{\cfRendBALo}{0.0001}
\newcommand{\cfRendBAHi}{0.0005}
\newcommand{\seedN}{5}

\newcommand{\seedBmin}{0.1226}
\newcommand{\seedBmax}{0.1228}

\newcommand{\synLowDimErr}{0.5}
\newcommand{\synTMVerr}{1.9}
\newcommand{\synTMerr}{0.9}
\newcommand{\synXerr}{25}
\newcommand{\synGBexcess}{0.5}

\newcommand{\synFloorX}{0.312}

\newcommand{\synMlpX}{0.318}
\newcommand{\synGBX}{0.449}

\newcommand{\cfsProbeArchPhi}{0.860}
\newcommand{\cfsProbeArchPhiM}{0.993}

\newcommand{\epOneExMin}{-0.2}
\newcommand{\epOneExMax}{+0.5}
\newcommand{\epOneShareMin}{70}
\newcommand{\epOneShareMax}{72}
\newcommand{\epTwoExMin}{-0.3}
\newcommand{\epTwoExMax}{+0.6}
\newcommand{\epTwoShareMin}{45}
\newcommand{\epTwoShareMax}{49}
\newcommand{\epThreeExMin}{-0.9}
\newcommand{\epThreeExMax}{+1.8}
\newcommand{\epThreeShareMin}{24}
\newcommand{\epThreeShareMax}{26}

\newcommand{\epTopNMin}{3{,}682}
\newcommand{\epTopNMax}{7{,}872}
\newcommand{\epMlpLowMax}{1.7}
\newcommand{\epMlpThirdMin}{2.1}
\newcommand{\epMlpThirdMax}{5.2}
\newcommand{\epShareSpreadMax}{4}

\newcommand{\epQStEgPct}{0.2}

\newcommand{\epQStEngVar}{8.8}
\newcommand{\epQStCnnVar}{10.5}
\newcommand{\epQStCnnOverEng}{0.0009}

\newcommand{\epEStEgPct}{0.2}

\newcommand{\epEStEngVar}{7.7}
\newcommand{\epEStCnnVar}{9.3}
\newcommand{\epEStCnnOverEng}{0.0008}

\newcommand{\epSStEgPct}{0.2}

\newcommand{\epSStEngVar}{8.0}
\newcommand{\epSStCnnVar}{9.6}
\newcommand{\epSStCnnOverEng}{0.0009}

\newcommand{\epLadNullMax}{0.0001}
\newcommand{\epLadZenMin}{0.0250}
\newcommand{\epLadZenMax}{0.0265}
\newcommand{\epLadMuMin}{0.0259}
\newcommand{\epLadMuMax}{0.0264}
\newcommand{\epRatioResolved}{2}
\newcommand{\epSampleFracDiffMax}{0.0003}
\newcommand{\epSampleMedDiffMax}{0.0023}

\newcommand{\epStCnnVarHiMin}{7.8}
\newcommand{\epStCnnVarHiMax}{9.8}
\newcommand{\epStEngVarHiMin}{4.4}
\newcommand{\epStEngVarHiMax}{5.1}
\newcommand{\epStMuPctHiMin}{1.2}
\newcommand{\epStMuPctHiMax}{2.3}
\newcommand{\epStEgPctHiMin}{0.6}
\newcommand{\epStEgPctHiMax}{1.0}

\newcommand{\epAbsImg}{0.1431}
\newcommand{\epCoreImg}{0.1432}
\newcommand{\epAbsRes}{0.1423}
\newcommand{\epCoreRes}{0.1425}
\newcommand{\epAbsVar}{11.6}

\newcommand{\epKcdcExOneMin}{6}
\newcommand{\epKcdcExOneMax}{7}
\newcommand{\epSpreadGbOneMin}{0.075}
\newcommand{\epSpreadGbOneMax}{0.085}
\newcommand{\epSpreadKcOneMin}{0.057}
\newcommand{\epSpreadKcOneMax}{0.070}
\newcommand{\epSpreadOrOneMin}{0.069}
\newcommand{\epSpreadOrOneMax}{0.081}
\newcommand{\epKcdcExTwoMin}{15}
\newcommand{\epKcdcExTwoMax}{16}
\newcommand{\epSpreadGbTwoMin}{0.011}
\newcommand{\epSpreadGbTwoMax}{0.017}
\newcommand{\epSpreadKcTwoMin}{0.023}
\newcommand{\epSpreadKcTwoMax}{0.036}
\newcommand{\epSpreadOrTwoMin}{0.049}
\newcommand{\epSpreadOrTwoMax}{0.060}
\newcommand{\epKcdcExThreeMin}{28}
\newcommand{\epKcdcExThreeMax}{32}

\newcommand{\epKcdcBiasMin}{+0.010}
\newcommand{\epKcdcBiasMax}{+0.055}
\newcommand{\epThrData}{0.15}
\newcommand{\epThrProtonMin}{38}
\newcommand{\epThrProtonMax}{40}
\newcommand{\epFrShareOneMin}{73}
\newcommand{\epFrShareOneMax}{75}

\newcommand{\epFrInflTwoMin}{23}
\newcommand{\epFrInflTwoMax}{25}
\newcommand{\epFrShareThreeMin}{26}
\newcommand{\epFrShareThreeMax}{28}
\newcommand{\epFrInflThreeMin}{12}
\newcommand{\epFrInflThreeMax}{15}

\newcommand{\epFrKcdcOverMin}{12}
\newcommand{\epFrKcdcOverMax}{58}
\newcommand{\epFrFloorExMin}{0.7}
\newcommand{\epFrFloorExMax}{0.8}
\newcommand{\epFrExCiLo}{0.4}
\newcommand{\epFrExCiHi}{1.1}
\newcommand{\epFrFloorMin}{0.1449}
\newcommand{\epFrFloorMax}{0.1465}
\newcommand{\epFrGenThreeMin}{0.1194}
\newcommand{\epFrGenThreeMax}{0.1261}
\newcommand{\epFrRawThreeMin}{0.0939}
\newcommand{\epFrRawThreeMax}{0.1105}

\newcommand{\epWithinN}{45}
\newcommand{\epSpreadThreeKcSmallerN}{2}
\newcommand{\epMeasWNeMin}{0.015}
\newcommand{\epMeasWNeMax}{0.065}
\newcommand{\epMeasWAgeMin}{0.007}
\newcommand{\epMeasWAgeMax}{0.093}
\newcommand{\epMeasWZeMin}{0.1}
\newcommand{\epMeasWZeMax}{1.2}
\newcommand{\epSelAcc}{49.1}
\newcommand{\epSelAll}{0.1021}
\newcommand{\epSelAccS}{0.1115}

\newcommand{\epSelDiff}{-0.00004}
\newcommand{\epSelDiffLo}{-0.00025}
\newcommand{\epSelDiffHi}{+0.00014}
\newcommand{\epSelRej}{0.0923}
\newcommand{\epSelFloorAcc}{0.1106}
\newcommand{\epSelFloorRej}{0.0912}
\newcommand{\epSelFloorAccLo}{0.1084}
\newcommand{\epSelFloorAccHi}{0.1121}
\newcommand{\epSelFloorRejLo}{0.0898}
\newcommand{\epSelFloorRejHi}{0.0927}
\newcommand{\epSelSpreadAll}{0.060}
\newcommand{\epSelSpreadAccAll}{0.112}

\newcommand{\epKSelRetrainMax}{0.1}
\newcommand{\epBenchRefLowMin}{+0.1}
\newcommand{\epBenchRefLowMax}{+0.6}
\newcommand{\epBenchRefThirdMin}{+1.7}
\newcommand{\epBenchRefThirdMax}{+2.4}
\newcommand{\epBenchRankThirdMax}{+8}

\newcommand{\epBenchCovMin}{60}
\newcommand{\epBenchCovMax}{95}
\newcommand{\epBenchReps}{20}

\newcommand{\epBenchLowAllMin}{0.0}
\newcommand{\epBenchLowAllMax}{+0.6}
\newcommand{\epBenchImpliedMin}{0.0}
\newcommand{\epBenchImpliedMax}{+1.4}
\newcommand{\epSensLowAbsMax}{2.2}
\newcommand{\epSensThirdLo}{-8.1}
\newcommand{\epSensThirdHi}{+3.9}
\newcommand{\epSensRefAbsMax}{1.8}
\newcommand{\epSensCiExcl}{2}
\newcommand{\epSensN}{12}
\newcommand{\epSensCiExclMax}{0.1}
\newcommand{\epSensSetWider}{9}
\newcommand{\epSensThirdNoRankLo}{-2.4}
\newcommand{\epSensThirdNoRankHi}{+3.8}
\newcommand{\epSensSixMaxPct}{1.3}
\newcommand{\epSensNnMin}{1}
\newcommand{\epSensNnMax}{17}
\newcommand{\epSensManhattanMax}{3.2}
\newcommand{\epClosureIncData}{0.280}
\newcommand{\epClosureIncSim}{0.282}
\newcommand{\epClosureMaxDiff}{0.017}
\newcommand{\epClosureZMax}{0.02}

\newcommand{\epMeasuredN}{13{,}848{,}509}
\newcommand{\epMeasuredTotal}{389{,}893{,}448}
\newcommand{\epMeasuredKeep}{5.1}
\newcommand{\epNeCut}{4.0}

\newcommand{\epMeasClosestOne}{proton}
\newcommand{\epMeasClosestTwo}{proton}
\newcommand{\epMeasClosestThree}{the mixture}

\newcommand{\epMeasZLowMin}{0.04}
\newcommand{\epMeasZLowMax}{0.32}
\newcommand{\epMeasZHiMin}{0.42}
\newcommand{\epMeasZHiMax}{0.54}

\newcommand{\epMeasZHiN}{197}
\newcommand{\epMeasZMixMin}{0.42}
\newcommand{\epMeasZMixMax}{0.55}
\newcommand{\epMeasIncDataMin}{0.276}
\newcommand{\epMeasIncDataMax}{0.283}
\newcommand{\epMeasIncSimMin}{0.282}
\newcommand{\epMeasIncSimMax}{0.287}
\newcommand{\epMeasIncDiffMax}{0.008}

\newcommand{\toyNoise}{0.30}
\newcommand{\toyGamma}{0.300}
\newcommand{\toyGammaLo}{0.297}
\newcommand{\toyGammaHi}{0.304}
\newcommand{\toyLin}{0.523}
\newcommand{\toyGB}{0.307}
\newcommand{\toyLinEx}{74}
\newcommand{\toyGBEx}{+2.5}
\newcommand{\toyGBExAbs}{2.5}
\newcommand{\toyNmax}{64{,}000}

\newcommand{\calibCovMin}{77}
\newcommand{\calibCovMax}{89}
\newcommand{\calibLowMin}{33}
\newcommand{\calibLowMax}{70}
\newcommand{\calibRestMin}{83}
\newcommand{\calibRestMax}{100}
\newcommand{\calibNfail}{4}

\newcommand{\cbTopShowersMin}{1{,}142}
\newcommand{\cbTopShowersMax}{2{,}487}
\newcommand{\cbKaTopShowers}{4{,}068}
\newcommand{\cbOverlapMin}{93.5}
\newcommand{\cbOverlapMax}{93.6}

\newcommand{\cbThreeShowersEpos}{11{,}288}
\newcommand{\cbThreeShowersSib}{11{,}602}
\newcommand{\cbTwoExMin}{+0.7}
\newcommand{\cbTwoExMax}{+1.9}
\newcommand{\cbTwoShareMin}{55}
\newcommand{\cbTwoShareMax}{57}
\newcommand{\cbTwoKaShareMin}{45}
\newcommand{\cbTwoKaShareMax}{49}
\newcommand{\cbTwoKcalOverMin}{53}
\newcommand{\cbTwoKcalOverMax}{54}

\newcommand{\cbTwoKcalSpreadMax}{0.17}
\newcommand{\cbTwoKaKcalSpreadMax}{0.04}

\newcommand{\cbTwoSensLo}{-0.9}
\newcommand{\cbTwoSensHi}{+1.7}

\newcommand{\cbThreeExMin}{+1.9}
\newcommand{\cbThreeExMax}{+2.9}
\newcommand{\cbThreeShareMin}{17}
\newcommand{\cbThreeShareMax}{21}
\newcommand{\cbThreeKaShareMin}{24}
\newcommand{\cbThreeKaShareMax}{26}
\newcommand{\cbThreeKcalOverMin}{100}
\newcommand{\cbThreeKcalOverMax}{111}

\newcommand{\cbThreeMLPdiffMax}{2.6}
\newcommand{\cbThreeSensLo}{-4.5}
\newcommand{\cbThreeSensHi}{+5.1}

\newcommand{\cbWithinMin}{1.4}
\newcommand{\cbWithinMax}{2.2}
\newcommand{\cbBetweenMin}{15}
\newcommand{\cbBetweenMax}{22}
\newcommand{\cbSibNsettings}{two settings}

\newcommand{\cbSibSensThreeHi}{+5.05}
\newcommand{\cbNcrit}{12}
\newcommand{\cbNcritMet}{11}

\newcommand{\keyOneMergedPct}{55}
\newcommand{\keyTwoMergedPct}{12}
\newcommand{\keyThreeMergedPct}{2}
\newcommand{\keyFloorMax}{1}
\newcommand{\keyExMax}{0.8}

\newcommand{\keyTwoEx}{+1.4}
\newcommand{\keyTwoExLo}{+0.6}
\newcommand{\keyTwoExHi}{+2.3}

\newcommand{\keyCombMax}{6}

\newcommand{\btDeadSeeds}{3}
\newcommand{\btGnnBest}{4}

\newcommand{\btGnnGainMax}{3.3}
\newcommand{\btGnnGainMin}{1.3}

\newcommand{\btMeanMax}{11.9}
\newcommand{\btMeanMin}{5.5}
\newcommand{\btPairsDistinct}{20}
\newcommand{\btPairsTotal}{36}

\newcommand{\btThreeBest}{12.3}
\newcommand{\btThreeEng}{9.0}
\newcommand{\btThreeEns}{6.9}
\newcommand{\btThreeEvents}{1{,}500{,}000}
\newcommand{\btThreeNull}{0.0}

\newcommand{\btThreeSigEns}{0.1437}
\newcommand{\btThreeSigNine}{0.1473}
\newcommand{\btThreeSigOracle}{0.0806}
\newcommand{\btTrainings}{120}

\renewcommand\paragraph[1]{\par\addvspace{1.2ex plus .4ex minus .2ex}\noindent{\normalfont\bfseries\itshape #1}\enskip\ignorespaces}

\hypersetup{colorlinks=true,linkcolor=black,citecolor=black,urlcolor=black,filecolor=black}
\begin{document}

\title[Is the learner or the summary the limit?]{Is the learner or the summary the limit? Auditing
machine-learning energy reconstruction for air-shower arrays on KASCADE and GRAPES-3}

\author[1]{\fnm{Mugdha Raje} \sur{Ashiya}\,\orcidlink{0009-0007-8652-9157}}

\author*[1,2]{\fnm{Samarjeet} \sur{Malik}\,\orcidlink{0009-0003-9924-7432}}\email{samarjeetmalik.gs5@gmail.com}

\affil[1]{\orgdiv{Department of Computer Science and Engineering}, \orgname{Indian Institute of
Technology Jodhpur}, \orgaddress{\city{Jodhpur}, \country{India}}}

\affil[2]{\orgdiv{Department of Computer Science}, \orgname{Ashoka University}, \orgaddress{\city{Sonipat}, \postcode{131029}, \state{Haryana}, \country{India}}}

\abstract{\unboldmath Air-shower arrays reduce hundreds of station signals to a few fitted shower parameters, the summary on which machine-learning energy estimators are usually trained. When such an estimator plateaus, either the model is too weak or the summary is exhausted; training diagnostics cannot tell which, though the remedies are opposite. We separate them with an audit built around the \emph{floor}, the lowest squared error reachable from the summary, estimated without training a model. It also checks whether each input was measured or only computed from the fit, which cannot add information. We apply it to public KASCADE simulations for three hadronic-interaction models, $\epMeasuredN$ measured KASCADE events and GRAPES-3 simulations. With the energy decade known, gradient boosting on nine reconstructed features comes within $\epSensLowAbsMax\%$ of the estimated floor below $10^{16}$\,eV under every estimator setting and model. The estimate is accurate to about 1\% in a semi-synthetic test but is not a bound; its intervals under-cover. Replacing reconstructed by true particle numbers removes $\epOneShareMin$--$\epOneShareMax\%$ of the error variance below $10^{15}$\,eV and $\epThreeShareMin$--$\epThreeShareMax\%$ at $10^{16}$--$10^{17}$\,eV, yet four station encoders recover only $\btMeanMin$--$\btMeanMax\%$ of that gap between $10^{14}$ and $10^{16}$\,eV. A pre-registered repeat on the KASCADE-Grande reconstruction meets $\cbNcritMet$ of $\cbNcrit$ model-by-criterion checks, failing one narrowly. On GRAPES-3, images rendered from the fit add nothing. In measured data the mean muon size lies within every model's proton-to-iron range and the muon channel behaves as in simulation; neither fixes the energy scale, and all energy-labelled results are simulated.}
\keywords{extensive air showers, energy reconstruction, machine learning, Bayes risk, KASCADE,
hadronic interaction models}

\maketitle

\section{Introduction}
\label{sec:intro}

\paragraph{The summary.} A ground-based air-shower array records the particle front of a cosmic-ray
shower at a few hundred detector stations. The energy of the primary particle is not measured
directly; it has to be inferred from how many particles reach the ground and how they are spread
out. The standard reconstruction fits a lateral distribution
function, in most arrays the Nishimura--Kamata--Greisen (NKG) function \citep{kamata,greisen}, to the
station signals and keeps a handful of its parameters: the electron size $\Ne$ (the number of
electrons at ground level), the shower age $s$ (a shape parameter of the lateral profile) and the
core position. Together with the arrival direction and a muon observable, these parameters form the
\emph{summary} of the event. It is the summary that is archived and distributed, and it is the input
on which learned estimators of the primary energy and mass are usually trained
\citep{kuznetsov,kuznetsov-jinst,lhaaso-ml}.

\paragraph{The problem.} An estimator trained on the summary cannot recover what the fit discarded.
When its resolution stops improving, either the estimator is limited, by capacity, data or tuning,
or the summary has no more information about the energy to give. From inside the training loop the two look the same: training converges, validation scores
are stable and learning curves flatten. Yet they call for opposite responses. A larger network, or
an image-like input rebuilt from the archived quantities, can help only in the first case. In the
second, the only remedy is information the summary does not contain, such as the station signals
themselves. Reconstruction from the station signals can do better than the summary
\citep{erdmann,auger-dnn,icetop,sdrn}, but it needs those signals, and for most archived data sets
they are no longer available. Knowing which case applies therefore decides whether effort should go
into models or into data.

\paragraph{What is missing.} Feature ablations and architecture comparisons show which inputs or
learners do best, but not how far the best learner is from what the summary allows. That needs a
reference that does not depend on any particular learner.

\begin{figure}[tb]
\centering
\includegraphics[width=\columnwidth]{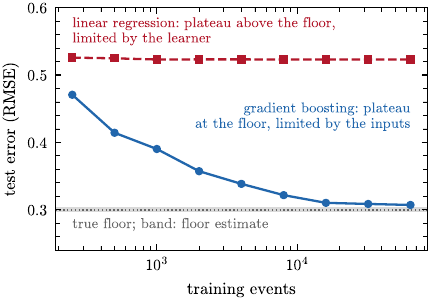}
\caption{Why a flattening learning curve is ambiguous (toy example of section~\ref{sec:method}). Both curves flatten at the largest training sets, but they flatten for different reasons. Linear regression stops far above the floor (dotted), limited by the learner; gradient boosting ends $\toyGBExAbs\%$ above it, so little is left for a better learner. The band is the floor estimate of section~\ref{sec:method}, which needs no trained model; for either curve on its own, only the floor says which case applies. Two features, target noise of standard deviation $\toyNoise$, so the true floor is $\toyNoise$}
\label{fig:toy}
\end{figure}

\paragraph{The idea.} Such a reference exists for squared loss. The lowest mean squared error that
any estimator can reach from the summary, its Bayes risk, is a property of the joint distribution of
summary and energy alone. It can be estimated without training a regressor, by asking how much the
energies of events that are close in the summary still differ (section~\ref{sec:method}). We call
its square root the \emph{floor} of the summary. A learner whose error sits at the estimated floor
is limited by the summary, as far as the estimate can tell; a learner well above it is limited by
itself (figure~\ref{fig:toy}). A second, exact fact
complements the estimate: an input computed deterministically from the summary, such as an image of
detector densities evaluated from the fitted lateral function, cannot lower the Bayes risk, however
the network that reads it is built.

\paragraph{Our approach.} We turn these two facts into a five-step audit (section~\ref{sec:method},
figure~\ref{fig:protocol}):
\begin{enumerate}
\item estimate the floor of the summary and compare trained estimators with it;
\item add the summary's quantities one at a time to see where the information enters;
\item check whether each input was measured or computed from the summary;
\item probe a trained model's output for its dependence on the summary, which needs no energy labels;
\item re-measure any inherited data selection where it is applied.
\end{enumerate}
Where the simulation provides them, an \emph{oracle} that replaces the reconstructed particle numbers by the true ones
adds a reference for what reconstruction loses. None of these components is new on its own; the
contribution is their combination into one procedure in which every estimated step is checked
against a control of known answer, and its application to two arrays. Only step~3 rests on an exact
guarantee.

\paragraph{Data.} The main testbed is the public KASCADE simulation \citep{antoni} released by the KASCADE
Cosmic-ray Data Centre (KCDC) \citep{kcdc,kcdcsim} for three hadronic-interaction models, QGSJET-II-04 \citep{qgsjet2}, EPOS-LHC \citep{eposlhc} and
SIBYLL-2.3d \citep{sibyll23d}, which differ notably in the muon content of showers; a conclusion that
holds for all three is not an artefact of one. The release includes the true particle numbers and
every station's signal, so the summary can be compared with what lies beyond it, and the measured
KASCADE events allow the label-free steps on real data. The second testbed is simulated GRAPES-3
events \citep{gupta}, for which only the summary is exported, so every input that can be built is a
function of it: a direct illustration of the exact guarantee.

\paragraph{Main results.} Table~\ref{tab:glance} collects the findings.
\begin{itemize}
\item \textbf{A larger model is not the main lever.} With the energy decade known, gradient boosting
on the nine reconstructed features is within $\epSensLowAbsMax\%$ of the estimated floor below
$10^{16}$\,eV, for every estimator setting and all three interaction models
(section~\ref{sec:kascade}).
\item \textbf{The cost of reconstruction falls with energy.} Replacing the reconstructed particle
numbers by the true ones removes $\epOneShareMin$--$\epOneShareMax\%$ of the error variance below
$10^{15}$\,eV but only $\epThreeShareMin$--$\epThreeShareMax\%$ between $10^{16}$ and $10^{17}$\,eV,
in the same way on all three models (section~\ref{sec:loss}).
\item \textbf{Station signals recover a small part of that cost.} Four station encoders, five seeds
each, close $\btMeanMin$--$\btMeanMax\%$ of the gap to the oracle on all three models between $10^{14}$ and
$10^{16}$\,eV (section~\ref{sec:stations}).
\item \textbf{A rendered image adds no information.} On GRAPES-3, networks given images rendered from
the fit do no better than the fit (section~\ref{sec:grapes}).
\item \textbf{Measured data do not contradict the simulation.} On measured KASCADE events, the
label-free checks agree with simulation; they do not test the energy scale
(section~\ref{sec:measured}).
\end{itemize}
Section~\ref{sec:method} defines the floor and the audit, section~\ref{sec:data} the data, and
sections~\ref{sec:kascade}--\ref{sec:grapes} apply the audit to KASCADE simulation, measured KASCADE
data and GRAPES-3. Section~\ref{sec:discussion} discusses what the results mean for arrays of this
kind and what they do not establish.

\begin{table}[tb]
\centering\footnotesize
\caption{Main results at a glance. Per-decade values are conditional on a known true-energy decade
(section~\ref{sec:fullrange} removes that condition), and energy-labelled results are simulated}
\label{tab:glance}
\begin{tabular}{@{}>{\raggedright\arraybackslash}p{0.34\columnwidth}>{\raggedright\arraybackslash}p{0.46\columnwidth}l@{}}
\toprule
Question & Result & Sec. \\
\midrule
Is the learner the limit below $10^{16}$\,eV? & Boosting within $\mathbf{\epSensLowAbsMax\%}$ of the estimated floor & \ref{sec:kascade} \\
How does the fixed KCDC formula compare? & Calibrated, $\mathbf{\epKcdcExOneMin}$--$\mathbf{\epKcdcExThreeMax\%}$ above the floor estimate & \ref{sec:kcdcest} \\
What do reconstructed sizes cost? & $\mathbf{\epOneShareMin}$--$\mathbf{\epOneShareMax\%}$ of error variance below $10^{15}$\,eV; $\mathbf{\epThreeShareMin}$--$\mathbf{\epThreeShareMax\%}$ at $10^{16}$--$10^{17}$\,eV & \ref{sec:loss} \\
How much do station networks recover? & $\mathbf{\btMeanMin}$--$\mathbf{\btMeanMax\%}$ of the oracle gap (four encoders, three models, $10^{14}$--$10^{16}$\,eV) & \ref{sec:stations} \\
Does it hold for another reconstruction? & KASCADE-Grande, three models: $\mathbf{\cbNcritMet}$ of $\cbNcrit$ pre-registered checks met & \ref{sec:comb} \\
Does a rendered image help on GRAPES-3? & No new information: CNN $\cfRendB$, boosting on the fit $\cfRendA$ & \ref{sec:grapes} \\
\bottomrule
\end{tabular}
\end{table}

\section{Method}
\label{sec:method}

The audit needs three things: a learner-independent reference for the best error the summary
allows, exact statements about which inputs can and cannot improve on it, and a procedure that
applies both with controls. This section gives them in that order.

\subsection{Floor, estimated floor and resolution}

Let $X$ be the station signals of an event, $T=t(X)$ the exported summary and $y=\lgE$ the
target. For squared loss the Bayes risk of predicting $y$ from $T$ is
\[
\Rstar(T)=\inf_f\E\bigl[(y-f(T))^2\bigr]=\E\bigl[\operatorname{Var}(y\mid T)\bigr],
\]
attained by $f(T)=\E[y\mid T]$. We call $\sigma^\star(T)=\sqrt{\Rstar(T)}$ the \emph{floor} of
$T$: the root-mean-square spread of the energy among events with identical summaries, which no
estimator can resolve because the summary does not record it. It is a property of the population and
is never observed. The \emph{estimated floor}
$\hat\sigma_{\mathrm B}(T)$ is a learner-independent estimate of it under the assumptions of the Gamma test (it trains no predictor, but depends on the metric, scaling and neighbour range), with an error of either sign; it is
not a lower bound, and an estimator may score below it. The \emph{resolution} $\sigma$ of a
trained estimator is the standard deviation of its residual on held-out events. Squared loss is
chosen because its Bayes predictor is a conditional expectation and risks decompose exactly
(appendix~\ref{app:proofs}). All statements below concern the conditional mean of the energy; a
channel that leaves it unchanged may still matter for composition or for calibrated
uncertainty.

\paragraph{The estimate.} The Gamma test \citep{gammatest,evansjones} estimates $\Rstar(T)$
without fitting a regressor. The idea: if two events have nearly the same summary, a smooth
regression function gives them nearly the same predicted energy, so whatever difference remains
between their true energies is the irreducible part. The test measures that difference for
neighbours at decreasing distance and extrapolates to zero distance. For each query event $i$ let $N_k(i)$ be its $k$-th nearest
neighbour in standardised feature space and
\begin{align*}
\gamma(k)&=\frac{1}{2n}\sum_{i=1}^{n}\bigl(y_i-y_{N_k(i)}\bigr)^2,\\
\delta(k)&=\frac{1}{n}\sum_{i=1}^{n}\bigl\|T_i-T_{N_k(i)}\bigr\|^2 ,
\end{align*}
with $n$ the number of query (test) events.
For a regression function with bounded derivatives $\gamma(k)\to\Rstar(T)$ as $\delta(k)\to0$,
and the intercept of a linear fit of $\gamma$ on $\delta$ over $k=1,\dots,10$ estimates it. Test
events are the queries, neighbours are drawn from other showers only, and intervals resample
showers. Before each use the estimator is run on the real feature matrix with a synthetic target
of known noise (a fixed smooth combination of the standardised electron size, muon size and age,
$0.9z_1+0.35z_2+0.08\tanh 2z_4$, scaled to $85\%$ of the target's spread, plus Gaussian noise of
standard deviation $0.10$ or $0.15$), and it must recover the noise level to within $10\%$ or the analysis stops. This control guards
against a failure of the estimator on the feature geometry; it is coarser than the precision needed,
which the sensitivity study of appendix~\ref{app:tables} and the benchmark address.
On a benchmark with known Bayes risks (appendix~\ref{app:synth}) the estimate is accurate to
$\synLowDimErr\%$ in up to six dimensions and to $\synTMVerr\%$ in eight, and it fails on a
$32$-dimensional measurement. It is therefore used here on summaries of at most nine features,
never on station signals.

\paragraph{A worked example.} Figure~\ref{fig:toy} shows the test on a toy problem with two standard
normal features and a target $y=x_1+0.3x_2^2+\varepsilon$, where $\varepsilon$ is Gaussian
noise of standard deviation $\toyNoise$, so the true floor is $\toyNoise$. The Gamma test, which trains
nothing, estimates $\toyGamma$ (central $95\%$ of $50$ independent estimates on samples of the same size:
$[\toyGammaLo,\toyGammaHi]$). With $\toyNmax$ training events,
gradient boosting reaches $\toyGB$, $\toyGBEx\%$ above the estimated floor, while linear
regression stops at $\toyLin$, $\toyLinEx\%$ above it. Both curves flatten at the largest training sets;
for either one on its own, only the floor says whether a better learner could help.

\subsection{Two propositions}

\begin{proposition}[A deterministic rendering cannot lower the Bayes risk]
\label{prop:bound}
Let $y$ be square-integrable and $g$ a deterministic measurable map. Then
$\inf_h\E[(y-h(g(T)))^2]\ge\inf_f\E[(y-f(T))^2]$, with equality if and only if
$\E[y\mid g(T)]=\E[y\mid T]$ almost surely.
\end{proposition}

An image of detector densities computed from the fitted parameters is such a $g(T)$. No encoder,
architecture or training budget moves the bound; only a separately measured channel can. The
statement concerns population risk, not how easily a model learns.

\begin{proposition}[What a measured channel can add]
\label{prop:channel}
If $y\in[a,b]$ almost surely and $M$ is any further observable, then
\begin{align*}
\Rstar(T)-\Rstar(T,M)&=\E\bigl[(\E[y\mid T,M]-\E[y\mid T])^2\bigr]\\
&\le\tfrac{(b-a)^2}{2}\,I(y;M\mid T).
\end{align*}
\end{proposition}

In words, a further channel lowers the Bayes risk by exactly the mean squared change it
causes in the best prediction, and that change is bounded by a multiple of the information the
channel carries about the energy beyond the summary. Both statements follow from the tower property; the inequality uses Pinsker's
inequality (appendix~\ref{app:proofs}). The converse of the second fails: a channel can carry information
about $y$ without moving its conditional mean, so a null result under squared loss concerns the
conditional mean only. The equality is what the ladder of section~\ref{sec:kascade} estimates rung by
rung; the bound is used only qualitatively, to say that a channel with no conditional information
cannot help.

\subsection{The audit}

\begin{figure*}[tb]
\centering
\includegraphics[width=0.780\textwidth]{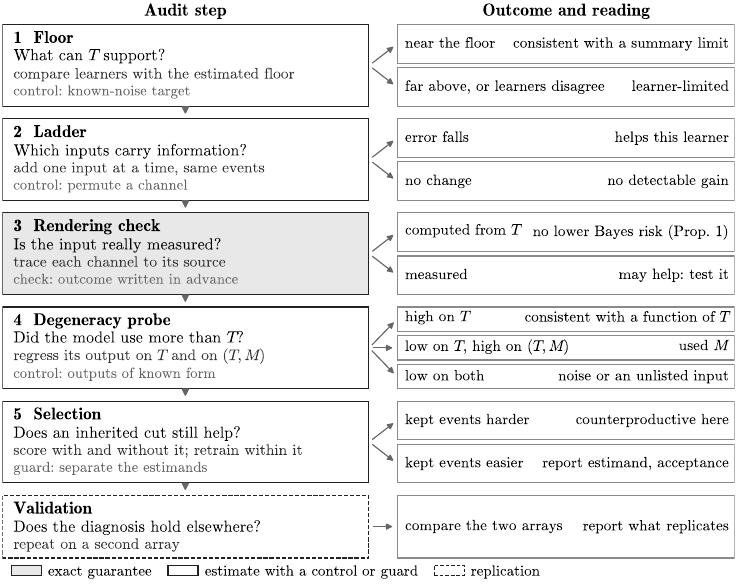}
\caption{The audit: five diagnostic steps and a validation on a second array with a different detector, for telling a representation-limited plateau from a learner-limited one. $T$ is the summary and $M$ a further measured channel. Step 3 rests on a guarantee (Proposition~\ref{prop:bound}). Steps 1, 2 and 4 are estimates with controls, step 5 is an estimate with a guard, and the validation is a replication}
\label{fig:protocol}
\end{figure*}

Figure~\ref{fig:protocol} gives the steps.
\begin{enumerate}
\item \textbf{Floor.} Compare estimators of different inductive bias
with the estimated floor: agreement at the floor reads as representation-limited for this
target and loss, as far as the estimate can tell; agreement well above it, or disagreement, as
learner-limited.
\item \textbf{Ladder.} Build the summary one quantity at a time from the fit outwards and score
each rung on the same events with paired intervals. A rung that lowers one learner's error helps
that learner; it is evidence that the Bayes risk fell only when the new error is below the
estimated floor of the previous rung.
\item \textbf{Rendering check.} For any input presented as richer than the summary,
establish whether it was measured or computed from the summary; if computed,
Proposition~\ref{prop:bound} applies.
\item \textbf{Degeneracy probe.} In the population, an output that is a function of the summary
alone is predicted perfectly by the summary, while an output that also used another channel is not;
with a finite regressor both score below one, so the probe is read against controls. Probe a trained model's output $m$ with
\[
R^2_{\mathrm{deg}}(T)=1-\frac{\E[(m-\E[m\mid T])^2]}{\operatorname{Var}(m)},
\]
the squared correlation ratio \citep{renyi}, estimated out of fold (five folds, grouped by shower in
simulation) with gradient boosting, on
$T$ alone and on $T$ with the other measured channels $M$. A model that used $M$ scores low on $T$
and near one on $(T,M)$; noise-like variation lowers both. Its scale is set by controls of known
construction (added noise, mixtures with a function of $M$), not by the value one: because
$\E[m\mid T]$ is approximated by a regressor of finite capacity, the probe measures what that
regressor can explain, and a function of $T$ scores somewhat below one. We call
$R^2_{\mathrm{deg}}(T,M)-R^2_{\mathrm{deg}}(T)$ the probe increment of $M$. The probe
needs only predictions and the listed inputs, so it applies to measured data.
\item \textbf{Selection.} Re-measure any
inherited selection in the regime where it is applied: score with and without it, retrain within it,
and report the quantity it was meant to protect, keeping three estimands apart: the operational
change, in which the scored population changes with the cut; the effect of training within the cut on
a fixed population; and the difficulty of the kept and the rejected events, from their floor
estimates.
\end{enumerate}
\begin{center}
\fbox{\parbox{\dimexpr\columnwidth-2\fboxsep-2\fboxrule\relax}{\small
\textbf{The probe in plain words.} The probe asks whether a trained model's output could have been
computed from the summary alone. A flexible regressor learns to predict the model's output from the
summary, and $R^2_{\mathrm{deg}}(T)$ is the share of the output's variance it explains. A model that
used only the summary scores close to one; a model that also relied on another channel $M$ scores
lower, but close to one again once $M$ is added. Because the regressor is finite, even a pure
function of the summary scores slightly below one, so values are read against controls with known
answers, not against one. The probe needs no energy labels, which is why it also runs on measured
data.}}
\end{center}

\paragraph{Oracle.} Where the simulation supplies true
particle numbers, replace the reconstructed ones by them: the change in
$\sigma$, with gradient boosting trained on the oracle features, is what observation and reconstruction
lose in the total sizes. It is a practical reference, not a strict bound for what a representation
of the station signals could recover, since their spatial pattern carries information beyond the
totals.

\paragraph{Headline and exploratory comparisons.} The comparisons the audit is built around are
gradient boosting against the floor estimate in each sample, the oracle-referenced gap, and the
rendering outcomes written down in advance (section~\ref{sec:grapes}). The ladder rungs, the
station representations, the encoder variants and the selection variants are exploratory: we give
paired shower intervals for them without a correction for multiplicity, and read them as
descriptive.

\paragraph{Estimators.} On the summary we use ridge regression, a random forest, histogram gradient
boosting ($600$ iterations, learning rate $0.06$) and a multilayer perceptron (two hidden layers of
$64$, early stopping), with settings fixed in advance and not tuned on test events. On the station
signals we use a small convolutional network, a residual network, a station transformer and a graph
network (appendix~\ref{app:encoders}). Throughout, $\lg$ denotes $\log_{10}$, and the resolution
$\sigma$ is the standard deviation of the residual; where the root-mean-square error differs, both
are given. Intervals are $95\%$ bootstrap intervals over
showers; paired intervals resample the same showers for both members of a comparison and do not
include the variation between training runs, which is reported separately where it matters.

\paragraph{Use of an AI assistant.} An AI assistant (Claude, Anthropic) was used to write
parts of the analysis code and to draft and edit parts of the text. The authors checked the code
and its outputs, every number in the text is generated from the results files by the analysis
pipeline and verified against them by script, and the authors take full responsibility for the
content of this article.

\section{Data}
\label{sec:data}

\begin{figure*}[tb]
\centering
\includegraphics[width=0.640\textwidth]{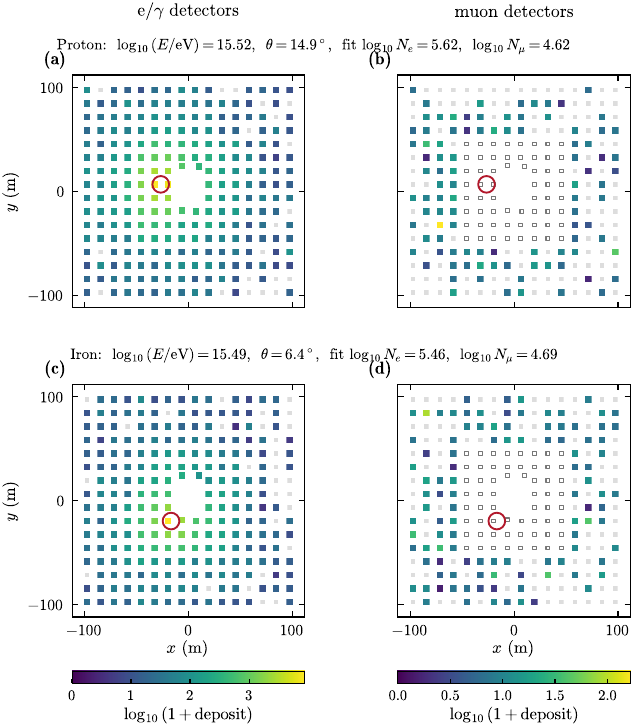}
\caption{What the fit keeps and what it discards: a proton ((a), (b); zenith angle $14.9^\circ$) and an iron nucleus ((c), (d); $6.4^\circ$) of nearly the same energy (QGSJET-II-04). Electromagnetic ((a), (c)) and muon ((b), (d)) deposits of the 252 stations (KCDC units), with the reconstructed core (circle). Grey squares: no signal; open squares in the muon panels: the 60 inner stations without a muon detector. The fit keeps the sizes, age and core and discards the station pattern}
\label{fig:events}
\end{figure*}

\begin{figure*}[tb]
\centering
\includegraphics[width=0.855\textwidth]{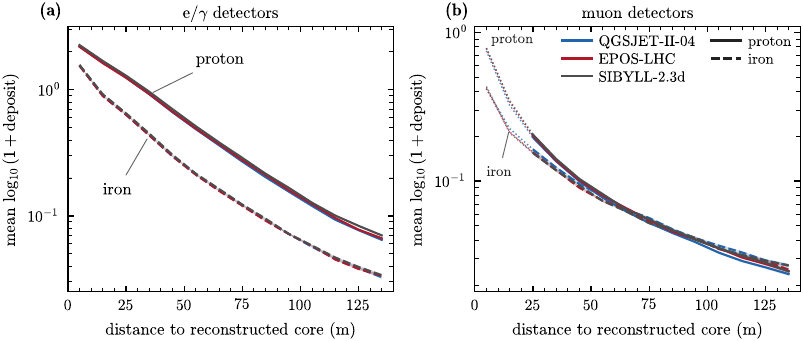}
\caption{Iron showers carry less electromagnetic signal than protons at every distance; protons carry more muon signal near the core, and beyond about $60$\,m the muon profiles are close. Mean $\log_{10}(1+\mathrm{deposit})$ in the electromagnetic (a) and muon (b) detectors against distance to the reconstructed core for protons (solid) and iron (dashed), true energy $10^{14.4}$--$10^{14.6}$\,eV, on the three interaction models. Muon means use the 192 stations with a muon detector; dotted segments mark bins populated by fewer than half of the events, because the inner stations have no muon detector. Bands: standard errors over events}
\label{fig:lateral}
\end{figure*}

\paragraph{KASCADE simulation.} KCDC releases simulations of the KASCADE array (252 stations on a
$200\times200$\,m$^2$ grid with separate electromagnetic and muon detectors \citep{antoni}) made
with CORSIKA \citep{corsika}, FLUKA below $200$\,GeV and one of three high-energy models
\citep{kcdcsim}: QGSJET-II-04, EPOS-LHC and SIBYLL-2.3d. Each set contains five primaries (proton,
helium, carbon, silicon, iron) generated on an $E^{-2}$ spectrum between $10^{14}$ and
$10^{18}$\,eV, several hundred thousand showers per primary, each thrown at three core positions; the
released events are those that pass KCDC's reconstruction and quality cuts. Each event carries the
reconstructed $\log_{10}\Ne$, $\log_{10}N_\mu$, age $s$, core position and zenith angle $\theta$, the
true energy, primary and particle numbers, and the per-station deposits and arrival times. The
nine features are the four fit parameters $(\log_{10}\Ne,s,X_c,Y_c)$, $\theta$ and $\cos\theta$, the
core distance, $\log_{10}N_\mu$ and the ratio $\log_{10}N_\mu-\log_{10}\Ne$; three of them are functions of
the others. Figure~\ref{fig:events} shows two events: the first proton and the first iron nucleus in the
QGSJET-II-04 file with true energy within $0.03$ of $\log_{10}(E/\mathrm{eV})=15.5$, zenith below
$15^\circ$ and core within $30$\,m. Figure~\ref{fig:lateral} shows the mean lateral profiles for
showers with zenith below $30^\circ$ and core within $60$\,m of the array centre. Its means include
stations without signal; beyond about $60$\,m iron is up to $14\%$ above proton in the muon quantity on
QGSJET-II-04. Its bands treat shower copies as independent, are at most about $1.7$ line widths wide,
and are not drawn on the dotted segments, where the error reaches about $3$ line widths.

\paragraph{Sampling and splits.} The release files are ordered by primary and energy, so a few
contiguous blocks are not a random sample of them; we found that such blocks change the oracle
of section~\ref{sec:loss} by more than its interval. The energy-resolved analysis draws events
uniformly at random (at most $300{,}000$ per decade); the station-level analysis, restricted to
$\lgE\in[14,15]$, reads $2{,}400$ evenly spaced blocks of $200$ events, whose primary fractions match
the full file to within $\epSampleFracDiffMax$ and whose median energy matches to within
$\epSampleMedDiffMax$ in $\lgE$. For $\lgE\in[15,16]$ up to $400{,}000$ events are drawn at random from
that window of each release. Because each shower is thrown three times, copies share
energy, primary and direction, and every split is by shower: $75\%$ of showers train and $25\%$
test. All scores average over events, so a shower counts once for each of its copies that passes the
release's cuts, which is the population an analysis of the release sees; floor estimates take neighbours from other showers only.
Showers are identified by their true energy and primary. That key is not unique in single
precision, so distinct showers can be grouped together (in the QGSJET-II-04 set, $\keyOneMergedPct\%$ of the
showers in the lowest decade share their key with another); appendix~\ref{app:showerkey} shows that a finer key changes no conclusion.

\paragraph{Measured KASCADE data.} KCDC also releases the measured events of the array with the
same reconstruction; section~\ref{sec:measured} uses them.

\paragraph{KASCADE-Grande.} KCDC's simulation of the larger KASCADE-Grande array, reconstructed with its COMBINED analysis, serves as a check on a second reconstruction between $10^{15}$ and $10^{17}$\,eV (section~\ref{sec:comb}).

\paragraph{GRAPES-3.} The second testbed is $\Nroot$ CORSIKA (QGSJET-II-04)/GEANT4 events of the GRAPES-3 array
at Ooty (about 400 scintillators at $8$\,m spacing and a $560$\,m$^2$ muon telescope read out as
$16$ modules \citep{gupta,hayashi,varsi-jinst}), simulated by the GRAPES-3 Collaboration for
proton, helium, nitrogen, aluminium and iron in one shower-size slice,
$\log_{10}\Ne\in[4.8,5.2]$, at $\theta\le24.6^\circ$. Only the summary and the $16$ module counts are
exported. The events come from $\nShowers$ showers ($\meanReuse$ copies on average); results are
given under a five-fold cross-fit by shower (appendix~\ref{app:grapes}).

\section{KASCADE: the summary across energy and interaction models}
\label{sec:kascade}

This section runs the audit on KASCADE simulation. It first asks whether standard learners have
reached the floor of the nine-feature summary (section~\ref{sec:plateau}) and how the
collaboration's own estimate compares (section~\ref{sec:kcdcest}). It then asks where in the summary
the information sits (the ladder), how much reconstruction costs (the oracle), whether the results
depend on knowing the energy decade, and how much of the cost the station signals recover.

\begin{table*}[tb]
\centering\small
\caption{Gradient boosting sits within about two per cent of the floor estimate in every interpreted decade, while the share of error variance tied to the reconstructed sizes falls from about $70\%$ to about $25\%$. The audit by decade of true primary energy on three hadronic-interaction models.
Events are a uniform random sample of each release (at most $300{,}000$ per decade), split
$75/25$ by CORSIKA shower. $\sigma$ is the standard deviation of the residual in $\lg E$
for gradient boosting on the nine reconstructed features (models trained within the decade); the floor estimate is the Gamma test with
test events as queries and neighbours from other showers; the excess is that of gradient
boosting over the floor estimate with a shower-bootstrap interval, which reflects shower sampling
only and under-covers (table~\ref{tab:floorbench}). The oracle replaces the
reconstructed electron and muon sizes by their true values; the last column is the
oracle-referenced error-variance gap $(\sigma^2-\sigma_{\rm oracle}^2)/\sigma^2$. The highest decade is
data-limited and is not interpreted. Bold: the oracle gap, the main result of this table, for the interpreted decades}
\label{tab:decades}
\footnotesize\setlength{\tabcolsep}{4pt}
\begin{tabular}{llrcccccc}
\toprule
$\lg(E/\mathrm{eV})$ & model & events & boosting & floor est. & excess & interval & oracle & oracle gap \\
\midrule
$14$--$15$ & QGSJET-II-04 & $300{,}000$ & $0.1481$ & $0.1479$ & $+0.1\%$ & $[-0.6,+0.8]$ & $0.0807$ & $\mathbf{70\%}$ \\
 & EPOS-LHC & $300{,}000$ & $0.1485$ & $0.1478$ & $+0.5\%$ & $[-0.2,+1.3]$ & $0.0783$ & $\mathbf{72\%}$ \\
 & SIBYLL-2.3d & $300{,}000$ & $0.1490$ & $0.1493$ & $-0.2\%$ & $[-0.9,+0.6]$ & $0.0805$ & $\mathbf{71\%}$ \\
\midrule
$15$--$16$ & QGSJET-II-04 & $300{,}000$ & $0.0869$ & $0.0864$ & $+0.6\%$ & $[-0.3,+1.4]$ & $0.0643$ & $\mathbf{45\%}$ \\
 & EPOS-LHC & $300{,}000$ & $0.0850$ & $0.0851$ & $-0.1\%$ & $[-0.9,+0.9]$ & $0.0607$ & $\mathbf{49\%}$ \\
 & SIBYLL-2.3d & $300{,}000$ & $0.0872$ & $0.0875$ & $-0.3\%$ & $[-1.1,+0.5]$ & $0.0639$ & $\mathbf{46\%}$ \\
\midrule
$16$--$17$ & QGSJET-II-04 & $78{,}023$ & $0.0614$ & $0.0603$ & $+1.8\%$ & $[-0.3,+4.3]$ & $0.0536$ & $\mathbf{24\%}$ \\
 & EPOS-LHC & $36{,}493$ & $0.0596$ & $0.0586$ & $+1.7\%$ & $[-1.6,+5.5]$ & $0.0521$ & $\mathbf{24\%}$ \\
 & SIBYLL-2.3d & $36{,}940$ & $0.0623$ & $0.0628$ & $-0.9\%$ & $[-3.7,+1.7]$ & $0.0535$ & $\mathbf{26\%}$ \\
\midrule
$17$--$18$ & QGSJET-II-04 & $7{,}872$ & $0.0564$ & $0.0481$ & $+17.2\%$ & $[+5.6,+33.7]$ & $0.0513$ & $17\%$ \\
 & EPOS-LHC & $3{,}682$ & $0.0604$ & $0.0482$ & $+25.3\%$ & $[+8.4,+54.5]$ & $0.0575$ & $9\%$ \\
 & SIBYLL-2.3d & $3{,}702$ & $0.0583$ & $0.0493$ & $+18.1\%$ & $[-0.8,+45.9]$ & $0.0539$ & $14\%$ \\
\bottomrule
\end{tabular}
\end{table*}

\subsection{A plateau at the estimated floor}
\label{sec:plateau}

\noindent\textbf{Key result.} Below $10^{16}$\,eV, boosting on the nine features is within $\epSensLowAbsMax\%$ of the estimated floor on all three models.

\begin{figure*}[tb]
\centering
\includegraphics[width=0.851\textwidth]{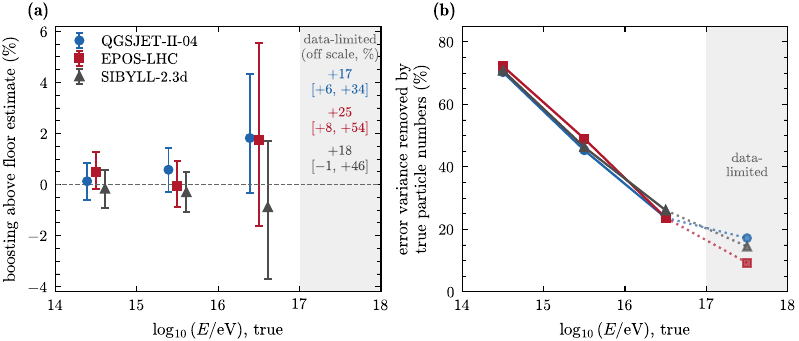}
\caption{KASCADE simulation by decade of true energy, for three hadronic-interaction models. (a) Gradient boosting on the nine reconstructed features above the Gamma-test floor estimate, with shower-bootstrap intervals, which reflect shower sampling only and under-cover; the highest decade (grey) is data-limited and off this scale, so its values and intervals are printed. (b) Oracle-referenced error-variance gap, the share of the summary's error variance removed by replacing reconstructed with true particle numbers. Its intervals are not shown; the dotted segment leads to the data-limited decade}
\label{fig:decades}
\end{figure*}

Table~\ref{tab:decades} and figure~\ref{fig:decades} give the audit by decade of true energy. Models
are trained and scored inside each decade, which assumes the decade is known;
section~\ref{sec:fullrange} removes that assumption. Each row reads in two steps. A small excess of boosting over the floor estimate is consistent with the learner having reached what the summary allows, as far as the
estimate can tell. The oracle gap then says how much of the remaining error would disappear with true
particle numbers.

We consider the three decades from $10^{14}$ to $10^{17}$\,eV. In the lowest, gradient boosting on the nine features differs from the floor estimate by $\epOneExMin$ to $\epOneExMax\%$; in the next, by $\epTwoExMin$ to $\epTwoExMax\%$. In the third it differs by $\epThreeExMin$ to $\epThreeExMax\%$. On this sample all nine intervals include zero (with a finer grouping of showers, the QGSJET-II-04 one between $10^{15}$ and $10^{16}$\,eV does not; appendix~\ref{app:showerkey}), but they reflect the sampling of showers only, ignore the bias of the estimator and, as shown below, are too narrow.

The other learners compare as follows. The multilayer perceptron is within $\epMlpLowMax\%$ of
gradient boosting in the two lower decades. In the third, which has fewer training events, it is
$\epMlpThirdMin$ to $\epMlpThirdMax\%$ above it. Ridge regression falls well short in every decade. The
known-noise control passed in every decade and model. The highest decade has only $\epTopNMin$ to
$\epTopNMax$ events per model, too few for either the learners or the floor estimate, so it is
reported but not interpreted.

\paragraph{How accurate is the floor estimate?} Two things matter: how much the estimate moves with
its settings (neighbour range, feature scaling, metric, coordinates), and how far it is from the
truth. Appendix~\ref{app:tables} gives the detail; the conclusions rest on the following results.

The settings matter more than sampling. In $\epSensSetWider$ of the $\epSensN$ cases they move the
estimate by more than its shower interval, and the Manhattan metric alone moves it by up to
$\epSensManhattanMax\%$. Even so, below $10^{16}$\,eV gradient boosting stays within
$\epSensLowAbsMax\%$ of the estimate under every setting.

To measure the distance from the truth, we keep the real nine features of each sample and replace
the energy by a semi-synthetic target whose floor is known. Below $10^{16}$\,eV the reference
estimate then reads high by only $\epBenchRefLowMin$ to $\epBenchRefLowMax\%$. Its shower intervals
ignore this small bias and cover the true floor in $\epBenchCovMin$ to $\epBenchCovMax\%$ of the
replicates instead of the nominal $95\%$. A pre-registered attempt to calibrate them failed on
held-out noise levels (appendix~\ref{app:tables}).

\paragraph{What survives the uncertainty.} Taking the settings, the measured bias and the
under-coverage together, three statements hold below $10^{16}$\,eV:
\begin{itemize}
\item gradient boosting is within $\epSensLowAbsMax\%$ of the estimate under every setting, and at
most $\epBenchImpliedMax\%$ above it once the estimate is corrected for the bias measured in the
benchmark;
\item the nonlinear estimators of different inductive bias agree with each other (in $\lgE\in[14,15]$
to within $\kfSpread$ in $\sigma$), while ridge regression falls well short;
\item hence more capacity on the nine features is not expected to buy more than about two per cent
in $\sigma$.
\end{itemize}
Two statements do not hold. Whether boosting lies above or below the true floor, and by how much
within that margin, is not determined. In the sparse decade $10^{16}$--$10^{17}$\,eV no comparison is
resolved better than the settings range. That range is $\epSensThirdNoRankLo$ to
$\epSensThirdNoRankHi\%$ even without the rank scaling, and the estimate there reads high by up to
$\epBenchRefThirdMax\%$. Throughout the paper, shower intervals quoted for the floor estimate reflect
the sampling of showers only and should be read as lower limits on its uncertainty.

\subsection{The standard reconstruction against the floor}
\label{sec:kcdcest}

\begin{table*}[tb]
\centering\footnotesize\setlength{\tabcolsep}{3.5pt}
\caption{KASCADE's own energy estimate $E_{\rm rec}$ (a fixed function of $\log_{10}\Ne$, $\log_{10}N_\mu$ and
zenith angle) against gradient boosting on the nine features and the floor estimate, on the same
test events. Raw: root-mean-square error of $\lg E_{\rm rec}$ with its mean bias; calibrated: boosting
on $E_{\rm rec}$ alone, which removes bias and nonlinearity (standard deviation of the residual). The last three columns give the spread
(largest minus smallest) of the mean bias across the five primaries for boosting, for the calibrated
KCDC estimate and for the oracle with true sizes. Bold: the lower resolution of boosting and the calibrated KCDC estimate, and the smaller bias spread of the two, whichever method that favours}
\label{tab:kcdcest}
\begin{tabular}{llccccccccc}
\toprule
 & & & & \multicolumn{3}{c}{KCDC estimate} & \multicolumn{3}{c}{bias spread across primaries} \\
$\lg(E/\mathrm{eV})$ & model & boosting & floor est. & raw (bias) & calibrated & above floor & boosting & KCDC & oracle \\
\midrule
$14$--$15$ & QGSJET-II-04 & $\mathbf{0.1481}$ & $0.1479$ & $0.1942$ ($+0.027$) & $0.1585$ & $+7\%$ & $0.075$ & $\mathbf{0.057}$ & $0.081$ \\
 & EPOS-LHC & $\mathbf{0.1485}$ & $0.1478$ & $0.1956$ ($+0.025$) & $0.1588$ & $+7\%$ & $0.083$ & $\mathbf{0.066}$ & $0.069$ \\
 & SIBYLL-2.3d & $\mathbf{0.1490}$ & $0.1493$ & $0.1972$ ($+0.034$) & $0.1586$ & $+6\%$ & $0.085$ & $\mathbf{0.070}$ & $0.073$ \\
\midrule
$15$--$16$ & QGSJET-II-04 & $\mathbf{0.0869}$ & $0.0864$ & $0.1216$ ($+0.010$) & $0.0994$ & $+15\%$ & $\mathbf{0.015}$ & $0.023$ & $0.060$ \\
 & EPOS-LHC & $\mathbf{0.0850}$ & $0.0851$ & $0.1229$ ($+0.022$) & $0.0984$ & $+16\%$ & $\mathbf{0.017}$ & $0.036$ & $0.049$ \\
 & SIBYLL-2.3d & $\mathbf{0.0872}$ & $0.0875$ & $0.1264$ ($+0.028$) & $0.1002$ & $+15\%$ & $\mathbf{0.011}$ & $0.035$ & $0.054$ \\
\midrule
$16$--$17$ & QGSJET-II-04 & $\mathbf{0.0614}$ & $0.0603$ & $0.0946$ ($+0.025$) & $0.0790$ & $+31\%$ & $0.030$ & $\mathbf{0.009}$ & $0.048$ \\
 & EPOS-LHC & $\mathbf{0.0596}$ & $0.0586$ & $0.1036$ ($+0.050$) & $0.0772$ & $+32\%$ & $\mathbf{0.021}$ & $0.024$ & $0.039$ \\
 & SIBYLL-2.3d & $\mathbf{0.0623}$ & $0.0628$ & $0.1089$ ($+0.055$) & $0.0801$ & $+28\%$ & $0.024$ & $\mathbf{0.018}$ & $0.042$ \\
\bottomrule
\end{tabular}
\end{table*}

\noindent\textbf{Key result.} Calibrated, the fixed KCDC formula is $\epKcdcExOneMin$ to $\epKcdcExThreeMax\%$ above the floor estimate, and the excess grows with energy.

\begin{figure*}[tb]
\centering
\includegraphics[width=0.843\textwidth]{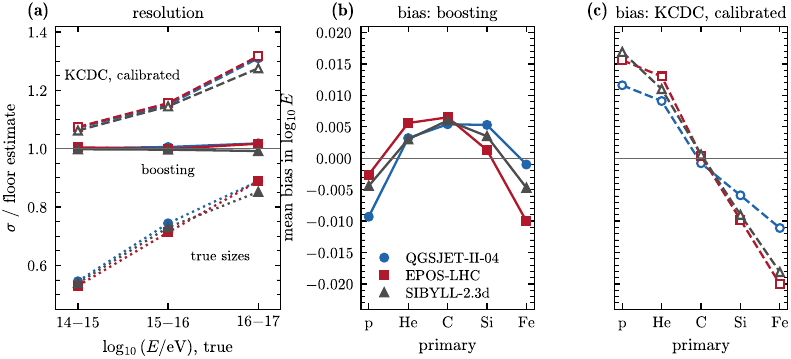}
\caption{(a) Resolution relative to the floor estimate for gradient boosting on the nine features (solid), the calibrated KCDC estimate (dashed, open markers) and the oracle with true sizes (dotted), on three interaction models. (b), (c) Mean bias in $\log_{10}E$ by primary between $10^{15}$ and $10^{16}$\,eV, for boosting (b) and for the calibrated KCDC estimate (c). Statistical intervals on the bias are not shown}
\label{fig:resolution}
\end{figure*}

KCDC releases with every event the collaboration's own energy estimate $E_{\rm rec}$, a fixed
function of the electron size, the muon size and the zenith angle. Table~\ref{tab:kcdcest} and
figure~\ref{fig:resolution} set it beside gradient boosting and the floor estimate on the same test
events. Used raw, it carries a mean bias of $\epKcdcBiasMin$ to $\epKcdcBiasMax$ in $\lg E$. To
compare resolutions fairly we calibrate it by fitting gradient boosting on $E_{\rm rec}$ alone, which
removes the bias and any nonlinearity. Like every model here, the calibration is fitted within the
decade.

The calibrated estimate falls further behind the floor as energy rises, on all three models. Below $10^{15}$\,eV it is $\epKcdcExOneMin$ to $\epKcdcExOneMax\%$ above the floor estimate. Between $10^{15}$ and $10^{16}$\,eV it is $\epKcdcExTwoMin$ to $\epKcdcExTwoMax\%$ above it. Between
$10^{16}$ and $10^{17}$\,eV the excess reaches $\epKcdcExThreeMin$ to $\epKcdcExThreeMax\%$.

This gap is not information the summary lacks. $E_{\rm rec}$ is itself a function of the summary, so
it cannot do better than the floor (Proposition~\ref{prop:bound}). The gap is what a fixed formula of
three quantities leaves unused of what the nine-feature summary allows, and a standard learner on the
same summary recovers it.

\paragraph{Bias by primary.} For an energy scale, the spread of the mean bias between primaries
matters as much as the resolution. Training for squared error on a mixture of primaries minimises the
average error and is free to trade bias between them, so a lower $\sigma$ need not mean less
dependence on composition.

Which estimator has the smaller spread changes with energy. Below $10^{15}$\,eV the bias spread of
gradient boosting ($\epSpreadGbOneMin$ to $\epSpreadGbOneMax$ in $\lg E$) is larger than that of the
calibrated KCDC estimate ($\epSpreadKcOneMin$ to $\epSpreadKcOneMax$). Between $10^{15}$ and $10^{16}$\,eV it is smaller: $\epSpreadGbTwoMin$ to $\epSpreadGbTwoMax$, against $\epSpreadKcTwoMin$ to $\epSpreadKcTwoMax$ for the calibrated KCDC estimate. Above, the calibrated KCDC estimate has the smaller spread on
$\epSpreadThreeKcSmallerN$ of the three models.

The spread is a coarse summary. Boosting's bias is not monotonic in mass
(figure~\ref{fig:resolution}b), because squared-error training pulls every primary towards the
mixture. That can bring the largest and smallest bias close together without making the estimate
independent of mass. Within each primary, boosting is the more precise of the two in all
$\epWithinN$ combinations of primary, decade and model. Where its bias spread is larger, it trades
composition dependence for resolution rather than being worse at both.

The composition dependence lies in the showers, not in their reconstruction. The oracle with true
particle numbers has a spread comparable to boosting's below $10^{15}$\,eV ($\epSpreadOrOneMin$ to
$\epSpreadOrOneMax$). Above $10^{15}$\,eV its spread is larger than either ($\epSpreadOrTwoMin$ to $\epSpreadOrTwoMax$ between $10^{15}$ and $10^{16}$\,eV). So the dependence comes from shower sizes that, at
fixed energy, depend on the primary mass. An analysis that needs an energy scale independent of
composition has to correct for mass, as KASCADE's unfolding does \citep{kascade2005}; a lower squared
error does not provide it.

\subsection{Where the information enters}

\begin{figure*}[tb]
\centering
\includegraphics[width=0.993\textwidth]{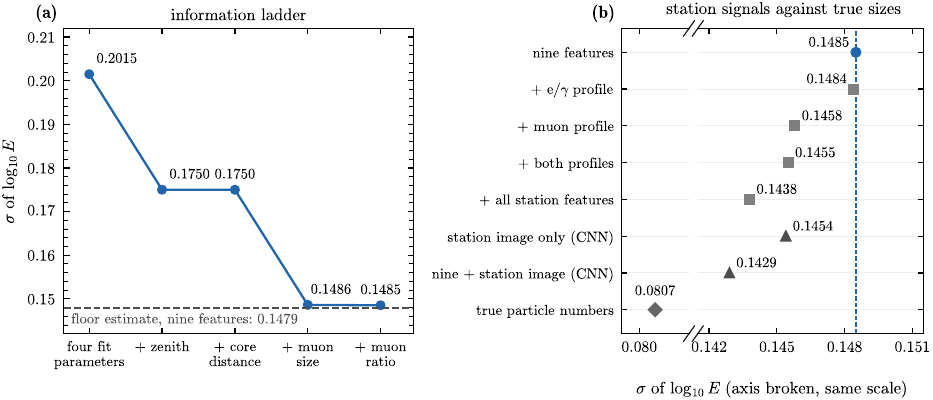}
\caption{The zenith angle and the muon size carry the information ladder (a); simulated station signals added to the summary improve on it but stay far from the oracle (b). KASCADE station-level sample, QGSJET-II-04, $10^{14}$--$10^{15}$\,eV. (a) One quantity added at a time from the fit outwards (``+ zenith'' adds $\theta$ and $\cos\theta$), with the floor estimate of the nine features. (b) The nine features (circle, dashed line); with hand-built station features added (squares); a three-seed CNN on the station image, alone and as a correction to the summary (triangles); the oracle with true sizes, which no detector measures (diamond): a reference for what the reconstructed totals cost, not a bound. Broken axis of constant scale}
\label{fig:kcdcaudit}
\end{figure*}

\noindent\textbf{Key result.} Two fitted quantities do most of the work: the zenith angle and the muon size.

The ladder adds the fitted quantities one at a time, from the fit outwards
(figure~\ref{fig:kcdcaudit}a; table~\ref{tab:ladderdec} in appendix~\ref{app:tables} for every
decade). Two steps account for most of the fall in $\sigma$, in every decade and on every model. In the station-level sample ($\lgE\in[14,15]$), adding the zenith angle lowers $\sigma$ by $\epLadZenMin$ to $\epLadZenMax$ across
the three models. Adding the muon size lowers it by a further $\epLadMuMin$ to $\epLadMuMax$.

The two other steps are the case Proposition~\ref{prop:bound} covers. The core distance and the muon ratio are
functions of quantities the learner already has, so they cannot lower the Bayes risk, and they change the learner's $\sigma$ by at most $\epLadNullMax$. A change large enough to resolve appears only for the ratio, on
$\epRatioResolved$ of the three models. It is a gain of the learner, not of information: boosting
finds the ratio easier to use than to build from its parts.

The ladder says which fitted quantities matter. The next question is how much reconstructing the particle numbers costs.

\subsection{What observation and reconstruction lose}
\label{sec:loss}

\noindent\textbf{Key result.} The share of error variance tied to the reconstructed sizes falls from $\epOneShareMin$--$\epOneShareMax\%$ to $\epThreeShareMin$--$\epThreeShareMax\%$ between the lowest and third decades.

To price the reconstruction, we replace the reconstructed electron and muon sizes by their true
values and train the same learner on them (the oracle column of table~\ref{tab:decades}). The share
of the summary's error variance this removes, $(\sigma^2-\sigma_{\rm oracle}^2)/\sigma^2$, is what we
call the \emph{oracle-referenced error-variance gap}. Because it compares two trained estimators, it
measures what the reconstructed totals cost this learner relative to true totals, not every piece of
information the detector loses.

The gap falls with energy (figure~\ref{fig:decades}b). It is $\epOneShareMin$ to
$\epOneShareMax\%$ below $10^{15}$\,eV and $\epTwoShareMin$ to $\epTwoShareMax\%$ between $10^{15}$
and $10^{16}$\,eV. Between $10^{16}$ and $10^{17}$\,eV it is $\epThreeShareMin$ to
$\epThreeShareMax\%$. In each decade the three interaction models agree to within
$\epShareSpreadMax$ percentage points. We do not interpret the data-limited top decade.

At low energy, then, most of the summary's error goes with the measured sizes. At higher energy most of the error remains even with true
sizes. This points to shower-to-shower fluctuations and the unknown primary mass, which no better
measurement of the totals removes. The oracle is a practical reference, not a bound: no detector
measures true particle numbers, and the station signals carry spatial and timing information beyond
the totals. Section~\ref{sec:stations} asks how much of that information they recover. First, section~\ref{sec:fullrange} asks which of these results survive when the energy decade is not known.

\subsection{Without knowing the decade}
\label{sec:fullrange}

\begin{table*}[tb]
\centering\footnotesize\setlength{\tabcolsep}{4pt}
\caption{The audit without knowing the true-energy decade. Models trained once on the whole release
($10^{14}$--$10^{18}$\,eV, up to $2\times10^6$ events, split by shower) and scored by decade of true
energy (root-mean-square error of $\lg E$), against boosting trained within each decade
(table~\ref{tab:decades}). ``Generated'': training on the simulated $E^{-2}$ spectrum; ``flat'':
training weighted to a flat distribution in $\lg E$. KCDC: its estimate calibrated by boosting on
$E_{\rm rec}$ alone (flat weights) and raw. Last two columns: share of error variance lost to
reconstruction (oracle with true sizes) with the decade known and with flat full-range training. Bold: the lowest error among the four estimators usable without knowing the decade (interpreted decades only)}
\label{tab:fullrange}
\begin{tabular}{llccccccc}
\toprule
 & & \multicolumn{3}{c}{boosting, nine features} & \multicolumn{2}{c}{KCDC estimate} & \multicolumn{2}{c}{lost to reconstruction} \\
$\lg(E/\mathrm{eV})$ & model & per decade & generated & flat & calibrated & raw & per decade & flat \\
\midrule
$14$--$15$ & QGSJET-II-04 & $0.1481$ & $\mathbf{0.1498}$ & $0.1626$ & $0.1768$ & $0.1945$ & $70\%$ & $73\%$ \\
 & EPOS-LHC & $0.1485$ & $\mathbf{0.1499}$ & $0.1628$ & $0.1762$ & $0.1958$ & $72\%$ & $75\%$ \\
 & SIBYLL-2.3d & $0.1490$ & $\mathbf{0.1511}$ & $0.1641$ & $0.1770$ & $0.1981$ & $71\%$ & $73\%$ \\
\midrule
$15$--$16$ & QGSJET-II-04 & $0.0869$ & $0.1218$ & $\mathbf{0.1086}$ & $0.1234$ & $0.1215$ & $45\%$ & $50\%$ \\
 & EPOS-LHC & $0.0850$ & $0.1172$ & $\mathbf{0.1044}$ & $0.1203$ & $0.1217$ & $49\%$ & $53\%$ \\
 & SIBYLL-2.3d & $0.0872$ & $0.1224$ & $\mathbf{0.1088}$ & $0.1245$ & $0.1269$ & $46\%$ & $52\%$ \\
\midrule
$16$--$17$ & QGSJET-II-04 & $0.0614$ & $0.1261$ & $\mathbf{0.0696}$ & $0.0871$ & $0.0939$ & $24\%$ & $26\%$ \\
 & EPOS-LHC & $0.0596$ & $0.1194$ & $\mathbf{0.0688}$ & $0.0842$ & $0.1028$ & $24\%$ & $26\%$ \\
 & SIBYLL-2.3d & $0.0623$ & $0.1221$ & $\mathbf{0.0699}$ & $0.0910$ & $0.1105$ & $26\%$ & $28\%$ \\
\midrule
$17$--$18$ & QGSJET-II-04 & $0.0564$ & $0.3182$ & $0.0653$ & $0.0784$ & $0.0826$ & $17\%$ & $18\%$ \\
 & EPOS-LHC & $0.0604$ & $0.3197$ & $0.0629$ & $0.0740$ & $0.1063$ & $9\%$ & $19\%$ \\
 & SIBYLL-2.3d & $0.0583$ & $0.3164$ & $0.0643$ & $0.0747$ & $0.0983$ & $14\%$ & $30\%$ \\
\bottomrule
\end{tabular}
\end{table*}

\noindent\textbf{Key result.} Without the decade the plateau and the energy dependence of the loss survive, but absolute resolutions are larger.

The analysis so far trains and scores inside a known decade of true energy, which an experiment
does not know. Every per-decade resolution and floor estimate above is conditioned on
$\lg(E/\mathrm{eV})\in[a,a+1]$, and this conditioning shrinks errors near the edges of the decade.
Table~\ref{tab:fullrange} removes it: the models are trained once on the whole release and scored by
decade of true energy.

How the training events are weighted then matters. Trained on the generated $E^{-2}$ spectrum,
boosting sees few high-energy showers and fails there. Between $10^{16}$ and $10^{17}$\,eV its error
is $\epFrGenThreeMin$ to $\epFrGenThreeMax$, worse than KCDC's raw formula ($\epFrRawThreeMin$ to
$\epFrRawThreeMax$). Weighted to a flat distribution in $\lg E$, it works, at a cost in resolution. Between $10^{15}$ and $10^{16}$\,eV it is $\epFrInflTwoMin$ to $\epFrInflTwoMax\%$ above the per-decade value; between $10^{16}$ and $10^{17}$\,eV, $\epFrInflThreeMin$ to $\epFrInflThreeMax\%$. The lowest decade also carries the lower edge of the simulation at
$10^{14}$\,eV.

Three conclusions survive without the decade:
\begin{itemize}
\item \emph{The plateau is not an artefact of the windows.} Over the whole range, boosting trained on
the generated spectrum is $\epFrFloorExMin$ to $\epFrFloorExMax\%$ above the floor estimate of the
nine features for the same population ($\epFrFloorMin$ to $\epFrFloorMax$). Joint shower intervals of
the excess lie within $\epFrExCiLo$ to $\epFrExCiHi\%$: measurably above the estimate, but by less
than one per cent.
\item \emph{The reconstruction share still falls with energy.} It falls from $\epFrShareOneMin$--$\epFrShareOneMax\%$ in the lowest decade to $\epFrShareThreeMin$--$\epFrShareThreeMax\%$ in the third (table~\ref{tab:fullrange}), each decade a few points above its known-decade value in section~\ref{sec:loss}.
\item \emph{A learner still beats the fixed formula.} KCDC's raw estimate remains $\epFrKcdcOverMin$
to $\epFrKcdcOverMax\%$ above flat-trained boosting in root-mean-square error between $10^{15}$ and
$10^{17}$\,eV.
\end{itemize}
The per-decade floors, and the comparison with the floor in section~\ref{sec:kcdcest}, are therefore
statements for a known decade. The absolute resolutions an experiment can expect from the summary
are those of the flat column of table~\ref{tab:fullrange}.

\subsection{What the station signals recover}
\label{sec:stations}

\begin{figure*}[tb]
\centering
\includegraphics[width=0.838\textwidth]{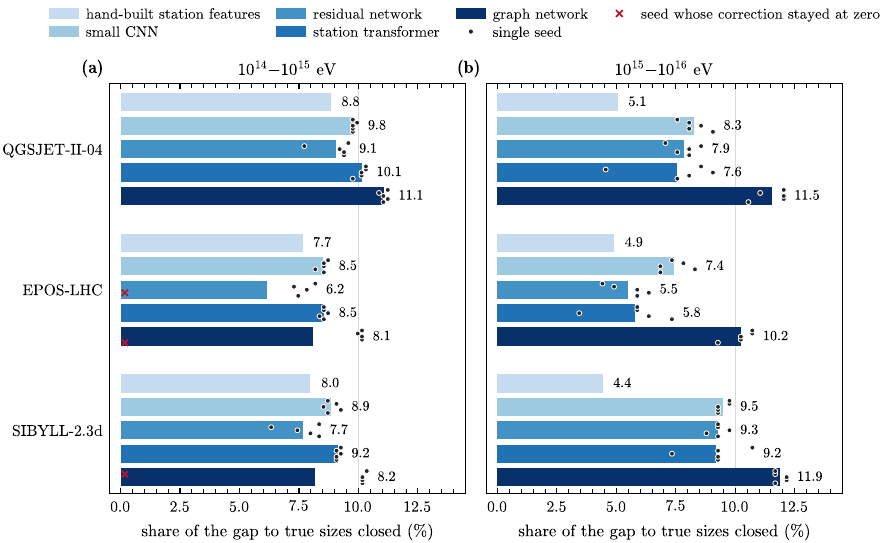}
\caption{Four station encoders close about a tenth of the variance gap between the nine-feature summary and the oracle, on every model and in both energy windows; the graph network leads in four of the six cells. (a) $10^{14}$--$10^{15}$\,eV, (b) $10^{15}$--$10^{16}$\,eV. Each encoder is trained as a correction to boosting on the nine features. Bars: means over five seeds; dots: single seeds, offset vertically so that equal values stay visible; crosses: seeds whose correction stayed at zero, counted as the pre-registration requires. The lightest bar is boosting with hand-built station features; the thin vertical line marks $10\%$. Encoder values as in table~\ref{tab:b2}, hand-built features as in table~\ref{tab:stations}; the seed-ensemble shares of table~\ref{tab:stations} are higher, because averaging the seeds' predictions reduces the error}
\label{fig:stations}
\end{figure*}

\begin{table*}[tb]
\centering\small
\caption{Station-level observables improve on the nine features on every model, but close only about a tenth of the gap to the oracle. What simulated station-level observables add, on three hadronic-interaction models
(for $\lg(E/\mathrm{eV})\in[14,15]$ $2{,}400$ evenly spaced blocks of $200$ events from each
release, about $410{,}000$ after the selection, for $[15,16]$ up to $400{,}000$ events drawn at random from that window; split by shower). Boosting and floor estimate on the nine reconstructed features; the
oracle with true sizes; boosting on the nine features plus all station features (radial
e/$\gamma$ and muon profiles, hit counts, totals and timing spread); and a convolutional network on
the simulated station image, trained as a correction to boosting (ensemble of three seeds). The last
columns give the share of the variance gap to the oracle closed by the station features and by the
network, and the network's change in $\sigma$ relative to the station features with a paired
shower-bootstrap interval. All $\sigma$ are standard deviations of the residual. Bold: the lowest $\sigma$ among inputs a detector can measure; its gain over the station features has a paired interval that excludes zero. Shares use $\sigma$ rounded to four decimals, so they carry up to about half a point of rounding error in the upper window}
\label{tab:stations}
\footnotesize\setlength{\tabcolsep}{4pt}
\setlength{\tabcolsep}{2.2pt}
\begin{tabular}{llrcccccccc}
\toprule
 & & & & floor & & + station & + image & \multicolumn{2}{c}{gap closed} & network $-$ features \\
$\lg(E/\mathrm{eV})$ & model & test & boosting & estimate & oracle & features & network & features & network & (paired interval) \\
\midrule
$14$--$15$ & QGSJET-II-04 & $102{,}483$ & $0.1485$ & $0.1479$ & $0.0807$ & $0.1438$ & $\mathbf{0.1429}$ & $8.8\%$ & $10.5\%$ & $-0.0009$ $[-0.0011,-0.0007]$ \\
 & EPOS-LHC & $102{,}629$ & $0.1484$ & $0.1477$ & $0.0772$ & $0.1442$ & $\mathbf{0.1433}$ & $7.7\%$ & $9.3\%$ & $-0.0008$ $[-0.0010,-0.0006]$ \\
 & SIBYLL-2.3d & $103{,}001$ & $0.1495$ & $0.1491$ & $0.0805$ & $0.1452$ & $\mathbf{0.1443}$ & $8.0\%$ & $9.6\%$ & $-0.0009$ $[-0.0011,-0.0007]$ \\
\midrule
$15$--$16$ & QGSJET-II-04 & $99{,}767$ & $0.0868$ & $0.0862$ & $0.0642$ & $0.0858$ & $\mathbf{0.0849}$ & $5.1\%$ & $9.6\%$ & $-0.0009$ $[-0.0010,-0.0008]$ \\
 & EPOS-LHC & $89{,}809$ & $0.0853$ & $0.0844$ & $0.0618$ & $0.0843$ & $\mathbf{0.0837}$ & $4.9\%$ & $7.8\%$ & $-0.0006$ $[-0.0007,-0.0005]$ \\
 & SIBYLL-2.3d & $92{,}905$ & $0.0872$ & $0.0865$ & $0.0638$ & $0.0863$ & $\mathbf{0.0852}$ & $4.4\%$ & $9.8\%$ & $-0.0011$ $[-0.0013,-0.0010]$ \\
\bottomrule
\end{tabular}
\end{table*}

\noindent\textbf{Key result.} Station signals improve on the summary, but four encoders on three models and in two energy windows close only $\btMeanMin$--$\btMeanMax\%$ of the gap to the oracle between $10^{14}$ and $10^{16}$\,eV. A larger pre-registered run is consistent with this but weak, because two of its three seeds failed to train.

The question here is how much of the oracle gap the simulated station-level observables can close
when they are added to the nine features. Table~\ref{tab:stations} answers
it as a share: $0\%$ would mean the station signals add nothing beyond the summary, $100\%$
that they match the oracle. The table's upper block is $\lgE\in[14,15]$, its lower block $[15,16]$.

\paragraph{Hand-built features.} We first give gradient boosting, with the nine features, radial
profiles of the electromagnetic and muon signals, hit counts, signal totals and the spread of arrival
times. In $\lgE\in[14,15]$ these close $\epQStEngVar\%$, $\epEStEngVar\%$ and $\epSStEngVar\%$ of the
variance gap on QGSJET-II-04, EPOS-LHC and SIBYLL-2.3d. Almost all of it comes from the muon profile,
whose shape differs between light and heavy primaries (figure~\ref{fig:lateral}) in a way the total
muon size does not capture. The electromagnetic profile adds $\epQStEgPct\%$, $\epEStEgPct\%$ and
$\epSStEgPct\%$ of the gap in $\sigma$.

\paragraph{Station image.} A convolutional network on the station image ($16\times16$ grid of
deposits and arrival times), trained as a correction to gradient boosting on the nine features,
closes $\epQStCnnVar\%$, $\epEStCnnVar\%$ and $\epSStCnnVar\%$ on the three models. That is better
than the hand-built profiles by $\epQStCnnOverEng$, $\epEStCnnOverEng$ and $\epSStCnnOverEng$ in
$\sigma$. The summary is therefore not predictively sufficient for the energy, and on every model the
measured signals recover about a tenth of what reconstruction loses.

\paragraph{One decade higher.} For $\lgE\in[15,16]$ the gap is smaller, and the network closes
$\epStCnnVarHiMin$ to $\epStCnnVarHiMax\%$ of it, so the fraction does not depend strongly on energy.
The hand-built features close less there ($\epStEngVarHiMin$ to $\epStEngVarHiMax\%$). The muon
profile adds less ($\epStMuPctHiMin$ to $\epStMuPctHiMax\%$ of the gap in $\sigma$) and the
electromagnetic profile somewhat more ($\epStEgPctHiMin$ to $\epStEgPctHiMax\%$).

\begin{table*}[tb]
\centering\small
\caption{Four station encoders on three interaction models and two energy windows, five seeds each.
Entries are the share of the variance gap to the oracle closed by each encoder (mean $\pm$ standard
deviation over seeds, in \%), each trained as a correction to boosting on the nine features; the
largest mean in each row (at the printed precision) is in bold. The last column applies the pre-registered test to the graph
network against the small CNN: the two differ only if their means are more than twice the standard
error of the difference apart (in brackets, graph network minus CNN in points). Encoder means span
$\btMeanMin$--$\btMeanMax\%$; $\btPairsDistinct$ of the $\btPairsTotal$ encoder pairs can be told apart
from seed noise}
\label{tab:b2}
\footnotesize\setlength{\tabcolsep}{4pt}
\begin{tabular}{llccccc}
\toprule
model & $\lg(E/\mathrm{eV})$ & small CNN & residual net & transformer & graph net & graph $\neq$ small? \\
\midrule
QGSJET-II-04 & $14$--$15$ & $9.8 \pm 0.1$ & $9.1 \pm 0.8$ & $10.1 \pm 0.2$ & $\mathbf{11.1 \pm 0.2}$ & yes ($+1.3$) \\
QGSJET-II-04 & $15$--$16$ & $8.3 \pm 0.6$ & $7.9 \pm 0.6$ & $7.6 \pm 1.8$ & $\mathbf{11.5 \pm 0.7}$ & yes ($+3.3$) \\
EPOS-LHC & $14$--$15$ & $\mathbf{8.5 \pm 0.2}$ & $6.2 \pm 3.5$ & $\mathbf{8.5 \pm 0.1}$ & $8.1 \pm 4.5$ & no ($-0.4$) \\
EPOS-LHC & $15$--$16$ & $7.4 \pm 0.6$ & $5.5 \pm 0.8$ & $5.8 \pm 1.4$ & $\mathbf{10.2 \pm 0.6}$ & yes ($+2.8$) \\
SIBYLL-2.3d & $14$--$15$ & $8.9 \pm 0.3$ & $7.7 \pm 0.8$ & $\mathbf{9.2 \pm 0.1}$ & $8.2 \pm 4.6$ & no ($-0.7$) \\
SIBYLL-2.3d & $15$--$16$ & $9.5 \pm 0.3$ & $9.3 \pm 0.3$ & $9.2 \pm 1.2$ & $\mathbf{11.9 \pm 0.3}$ & yes ($+2.4$) \\
\bottomrule
\end{tabular}
\end{table*}

\paragraph{Other encoders.} Larger and differently built encoders change this only a little. We trained four
of them on every model and in both windows, five seeds each, with one protocol fixed before training
(table~\ref{tab:b2} and figure~\ref{fig:stations}; the protocol is in appendix~\ref{app:encoders}): the CNN, a residual network about
thirty times its size, a transformer with one token per station, and a graph network over the
stations. Every encoder's mean share lies between $\btMeanMin$ and $\btMeanMax\%$, in all $24$
combinations. The pre-registered test of ``about a tenth'' asked for every mean to fall between $5$ and
$15\%$, so that description survives across encoders, models and energies. Within one model and window,
$\btPairsDistinct$ of the $\btPairsTotal$ encoder pairs differ by more than twice the standard error of
their difference. The graph network has the largest mean in $\btGnnBest$ of the six cells, and in those
four it beats the CNN by $\btGnnGainMin$ to $\btGnnGainMax$ points. The gain is largest between
$10^{15}$ and $10^{16}$\,eV, where $2.4$ to $3.3$ points is a quarter to two fifths of the CNN's own share. That
meets the pre-registered threshold for calling the advantage real (four cells of six). In the other
two cells, EPOS-LHC and SIBYLL-2.3d between $10^{14}$ and $10^{15}$\,eV, one of its five seeds failed
to train, and its spread is then too wide to tell it from the CNN.

\paragraph{Failed seeds.} Each network learns a correction to boosting and starts by outputting zero.
In $\btDeadSeeds$ of the $\btTrainings$ trainings of table~\ref{tab:b2} (one residual network and two
graph networks) the correction collapsed: the seed's $\sigma$ equals boosting's to within $10^{-5}$,
although early stopping kept epoch $6$, $13$ or $15$. We have not diagnosed the cause. As the
pre-registration requires, these seeds are counted rather than dropped. They lower the affected means
and widen their spreads.

\paragraph{Is the core position responsible?} Unlike the other encoders, the graph network is given
each station's position relative to the reconstructed core. We therefore trained it again, with the
same budget, on absolute station positions and without the core. It reaches $\epAbsImg$ on the image
alone (against $\epCoreImg$) and $\epAbsRes$ as a correction (against $\epCoreRes$). It then closes $\epAbsVar\%$ of the variance gap. This control shows that access to the reconstructed core does not explain the graph network's advantage.
Nor does its longer training in those runs: in table~\ref{tab:b2}, where every encoder had the same
$20$ epochs, the graph network still leads in four of six cells. A benefit from local message passing
between stations is consistent with this, although other differences in inductive bias cannot be
excluded.

\paragraph{More data and every input.} A natural objection is that the encoders are short of data
or of input. Before extracting any data for it, we pre-registered one larger run to test this. The
graph network was given station positions both relative to the core and absolute, together with the
nine features, with three seeds, on a new sample of $\btThreeEvents$ QGSJET-II-04 events with
$\lgE\in[14,15]$, about $3.7$ times the sample above (split by shower as before, so three quarters, less a validation tenth, are
used for training). The reading was fixed in advance: a share above $20\%$ would mean the summary
leaves substantial information that a stronger station model can use, and a share below $15\%$ would
be a strong negative. A control with the station images shuffled across events had to close less than
$1\%$ of the gap.

Formally the null criterion is met and the ensemble falls in the strong-negative band: the control closes $\btThreeNull\%$, and the three-seed ensemble
$\btThreeEns\%$ ($\sigma=\btThreeSigEns$, against $\btThreeSigNine$ for boosting and
$\btThreeSigOracle$ for the oracle on the same test events). Neither number carries the weight it was
meant to. Two of the three seeds collapsed to boosting's $\sigma$ in the way just described, and so did
the control, so its $0\%$ cannot be told apart from a network that did not train, and it does not show
that the pipeline is free of leakage. Averaging one trained correction with two zero corrections also
shrinks the ensemble's share to a little over half of the trained seed's. The informative result is
therefore that single seed. It closes $\btThreeBest\%$ of the gap, below $15\%$ and above boosting with
the hand-built station features ($\btThreeEng\%$). That is about one point more than the graph network
on the smaller sample, although the two use different test events. Its best validation epoch was $19$
of $20$, and the run used the training script's default of $20$ epochs rather than the $40$ of the
earlier graph-network runs (appendix~\ref{app:encoders}), which we noticed only afterwards. We read the
larger run as weak support for the B2 picture, not as the strong negative it was designed to give:
one trained seed, an uninformative control and a collapse rate of two in three, against three in
$\btTrainings$ in table~\ref{tab:b2}, which suggests a problem specific to this configuration.

\paragraph{What can be concluded.} Four encoder families, on three interaction models and in two
energy windows, agree that the station signals close about a tenth of the gap to the oracle
($\btMeanMin$--$\btMeanMax\%$), and the one trained seed of the larger run does not change that. This is
an empirical statement about these encoders and this training budget, not a bound.
What separates clearly from zero is each network's gain over the summary and over the hand-built
station features, for which table~\ref{tab:stations} gives paired shower intervals. (The CNN appears as
an ensemble of three seeds in table~\ref{tab:stations}, and as a mean of per-seed shares in
tables~\ref{tab:encoders} and~\ref{tab:b2}; the mean is lower, because averaging the seeds' predictions
reduces the error.) We cannot apportion the unrecovered remainder between two causes: the sparse
sampling of the shower by the array, which no function of the station signals undoes, and information
that a better model of those signals would extract.

\subsection{A second reconstruction: KASCADE-Grande}
\label{sec:comb}

\begin{table*}[tb]
\caption{The audit on the KASCADE-Grande COMBINED reconstruction, by decade of true energy and interaction
model, with the decade known; a uniform random subsample of at most 300{,}000 events per decade, split by
shower. $\sigma$ of $\hat y-y$ for $y=\lgE$ for boosting and the multilayer perceptron (MLP). Boosting's
excess over the floor estimate is in per cent, with a shower-bootstrap interval (a lower limit on the
uncertainty). The share is the fraction of boosting's error variance removed by true particle numbers. The
last column is the collaboration's COMBINED energy estimate $E$, calibrated on the training events. The
decade $10^{17}$--$10^{18}$\,eV holds only $\cbTopShowersMin$ to $\cbTopShowersMax$ distinct showers per
model and is not interpreted. Pre-registered criteria: the plateau (excess within $\pm2.2\%$ and $\pm5\%$, judged on the
point estimate), the falling share and the gain over $E$ are met on all three models; the plateau under every
estimator setting is met on QGSJET-II-04 and EPOS-LHC and not on SIBYLL-2.3d, where the excess over all
settings reaches $\cbSibSensThreeHi\%$ between $10^{16}$ and $10^{17}$\,eV}
\label{tab:comb}
\centering\footnotesize\setlength{\tabcolsep}{3.5pt}
\begin{tabular}{llrcccccccc}
\toprule
decade (eV) & model & events & boosting & MLP & floor est. & excess (\%) & interval & oracle & share & COMBINED $E$ \\
\midrule
$10^{15}$--$10^{16}$ & QGSJET-II-04 & 300{,}000 & 0.1036 & 0.1045 & 0.1028 & $+0.8$ & $[+0.1,+1.6]$ & 0.0690 & 56\% & 0.1591 \\
 & EPOS-LHC & 300{,}000 & 0.1017 & 0.1016 & 0.0998 & $+1.9$ & $[+0.9,+2.7]$ & 0.0671 & 57\% & 0.1560 \\
 & SIBYLL-2.3d & 300{,}000 & 0.1029 & 0.1031 & 0.1022 & $+0.7$ & $[-0.2,+1.4]$ & 0.0688 & 55\% & 0.1581 \\
\midrule
$10^{16}$--$10^{17}$ & QGSJET-II-04 & 124{,}604 & 0.0647 & 0.0646 & 0.0629 & $+2.9$ & $[+1.3,+4.4]$ & 0.0576 & 21\% & 0.1325 \\
 & EPOS-LHC & 57{,}857 & 0.0599 & 0.0601 & 0.0588 & $+1.9$ & $[-0.4,+4.4]$ & 0.0546 & 17\% & 0.1266 \\
 & SIBYLL-2.3d & 59{,}270 & 0.0656 & 0.0673 & 0.0637 & $+2.9$ & $[+0.5,+5.0]$ & 0.0594 & 18\% & 0.1309 \\
\bottomrule
\end{tabular}
\end{table*}

\noindent\textbf{Key result.} On the combined KASCADE-Grande reconstruction, with three interaction models, the plateau, the falling reconstruction share and the gain over the fixed formula meet their pre-registered criteria; the estimator-sensitivity criterion fails narrowly on one model.

KCDC also releases simulations reconstructed with the KASCADE-Grande COMBINED analysis, which uses a
larger array containing the KASCADE stations \citep{kcdcsim} and gives the same nine features. For
each interaction model, $\cbOverlapMin$ to $\cbOverlapMax\%$ of its showers also appear in that
model's KASCADE set, and none is shared between models, so this tests a different array and
reconstruction, not a different simulation. We ran the per-decade audit and the sensitivity scan on
QGSJET-II-04, EPOS-LHC and SIBYLL-2.3d unchanged. The first three criteria were fixed after counting showers per decade and before any training; the sensitivity criterion was added after the QGSJET-II-04 result and before its scan; all four were fixed before training on EPOS-LHC and SIBYLL-2.3d (\texttt{PREREG\_v3.md} in the code repository). A test above $10^{17}$\,eV was dropped at the start, for the reason given below. Of the $\cbNcrit$ model-by-criterion
checks, $\cbNcritMet$ are met (table~\ref{tab:comb}).

\emph{Plateau.} Boosting is $\cbTwoExMin$ to $\cbTwoExMax\%$ from the floor estimate between $10^{15}$
and $10^{16}$\,eV, inside the pre-set $\pm2.2\%$. Between $10^{16}$ and $10^{17}$\,eV it is
$\cbThreeExMin$ to $\cbThreeExMax\%$, inside the pre-set $\pm5\%$. Unlike on KASCADE, most intervals
exclude zero, so boosting may sit slightly above the estimate. The multilayer perceptron is within
$\cbThreeMLPdiffMax\%$ of boosting, and the known-noise control passed everywhere. Over all estimator
settings the lower decade stays within $\cbTwoSensLo$ to $\cbTwoSensHi\%$. In the sparser upper decade ($\cbThreeShowersEpos$ and $\cbThreeShowersSib$ showers on EPOS-LHC and SIBYLL-2.3d), \cbSibNsettings on SIBYLL-2.3d (five neighbours, two scalings) give $\cbSibSensThreeHi\%$, outside the limit by less than the rounding of the estimate, so that criterion is not met. Over all settings and models that decade spans $\cbThreeSensLo$ to $\cbThreeSensHi\%$, so a $\pm5\%$ test there has little power.

\emph{Reconstruction share and fixed formula.} On every model the share of error variance removed by true particle numbers falls from $\cbTwoShareMin$--$\cbTwoShareMax\%$ to $\cbThreeShareMin$--$\cbThreeShareMax\%$ between the two decades. On KASCADE the same decades give $\cbTwoKaShareMin$--$\cbTwoKaShareMax\%$ and $\cbThreeKaShareMin$--$\cbThreeKaShareMax\%$. The calibrated COMBINED energy estimate's $\sigma$ exceeds boosting's by $\cbTwoKcalOverMin$--$\cbTwoKcalOverMax\%$ and by $\cbThreeKcalOverMin$--$\cbThreeKcalOverMax\%$ in the two decades. Only part of that gap is composition dependence. Its mean bias still differs by up to $\cbTwoKcalSpreadMax$ in $\lg E$ between primaries, against at most $\cbTwoKaKcalSpreadMax$ for KCDC's KASCADE estimate, but that accounts for only $\cbBetweenMin$--$\cbBetweenMax\%$ of the variance gap; within each primary its $\sigma$ is still $\cbWithinMin$ to $\cbWithinMax$ times boosting's.

\emph{Above $10^{17}$\,eV.} The releases hold only $\cbTopShowersMin$ to $\cbTopShowersMax$ distinct
showers per model there, fewer than the $\cbKaTopShowers$ of the KASCADE decade already set aside, and
on all three models the floor estimate falls below the oracle, which no true floor can do. That
decade is run but not interpreted.

\section{Measured KASCADE data}
\label{sec:measured}

\noindent\textbf{Key result.} The mean muon size of the measured events lies inside every model's proton-to-iron bracket, and the muon channel is used as in simulation; neither check fixes the energy scale.

The results so far are simulated. The audit's conclusions are only as good as the simulations, so
we ask whether measured KASCADE events behave like them. Without true energies, steps 1, 2 and 5 and
the oracle cannot be run on measured events. Two checks can. First,
the distributions of $\log_{10}\Ne$, $\log_{10}N_\mu$, age and zenith angle in the data are compared with
each interaction model in decades of reconstructed energy, after reweighting the simulation from
its $E^{-2}$ spectrum to a knee spectrum (index $-2.7$ below $10^{15.6}$\,eV and $-3.1$ above
\citep{kascade2005}) for pure proton, pure iron and a five-species mixture of all released events (after the cuts and the knee weighting, each primary is $11$ to $30\%$ of it), and the mean
$\log_{10}N_\mu$ at fixed $\log_{10}\Ne$ is placed on each model's proton-to-iron scale. Second, estimators
trained on each simulation are applied to the data and probed (step 4): if an estimator depends on
the muon channel in the data as it does in simulation, the probe values agree. Both checks were
first run on pseudo-data drawn from half of the EPOS-LHC showers with a known spectrum and
composition: the pseudo-data are closest to EPOS-LHC in the muon-size distribution in every
decade, their mean $\log_{10}N_\mu$ sits on the EPOS-LHC proton-to-iron scale within
$\epClosureZMax$ of the mixture's value in every populated $\log_{10}\Ne$ bin, and the probe
increment of the muon channel is $\epClosureIncData$ on the pseudo-data against
$\epClosureIncSim$ on held-out EPOS-LHC simulation (largest difference over the three models
$\epClosureMaxDiff$).

\begin{table*}[tb]
\centering\footnotesize\setlength{\tabcolsep}{4pt}
\caption{In every decade and on every model the measured mean muon size lies between the proton and iron simulations. Measured KASCADE events against simulation, per decade of reconstructed energy, after the
common cuts (including $\log_{10}\Ne\ge4.0$). Mean $\log_{10}N_\mu$ of the data and of each model for pure
proton, the five-species mixture and pure iron, reweighted to a knee spectrum; and the
Wasserstein-1 distance between the data's $\log_{10}N_\mu$ distribution and each. Data are a random
$5.1\%$ of the release}
\label{tab:measured}
\begin{tabular}{lrlccccccc}
\toprule
 & & & \multicolumn{4}{c}{mean $\log_{10}N_\mu$} & \multicolumn{3}{c}{distance to data} \\
$\lg(E_{\rm rec}/\mathrm{eV})$ & events & model & data & proton & mixture & iron & proton & mixture & iron \\
\midrule
14--15 & $12{,}308{,}663$ & QGSJET-II-04 & $3.486$ & $3.456$ & $3.526$ & $3.623$ & $0.031$ & $0.040$ & $0.137$ \\
 &  & EPOS-LHC & $3.486$ & $3.459$ & $3.521$ & $3.611$ & $0.028$ & $0.034$ & $0.124$ \\
 &  & SIBYLL-2.3d & $3.486$ & $3.455$ & $3.522$ & $3.612$ & $0.031$ & $0.035$ & $0.126$ \\
\midrule
15--16 & $1{,}522{,}324$ & QGSJET-II-04 & $4.156$ & $4.126$ & $4.195$ & $4.248$ & $0.030$ & $0.039$ & $0.092$ \\
 &  & EPOS-LHC & $4.156$ & $4.135$ & $4.197$ & $4.247$ & $0.021$ & $0.041$ & $0.091$ \\
 &  & SIBYLL-2.3d & $4.156$ & $4.126$ & $4.193$ & $4.249$ & $0.030$ & $0.037$ & $0.093$ \\
\midrule
16--17 & $17{,}325$ & QGSJET-II-04 & $5.122$ & $5.037$ & $5.129$ & $5.214$ & $0.085$ & $0.019$ & $0.092$ \\
 &  & EPOS-LHC & $5.122$ & $5.057$ & $5.140$ & $5.224$ & $0.065$ & $0.022$ & $0.102$ \\
 &  & SIBYLL-2.3d & $5.122$ & $5.038$ & $5.129$ & $5.213$ & $0.084$ & $0.017$ & $0.091$ \\
\bottomrule
\end{tabular}
\end{table*}

\begin{figure*}[tb]
\centering
\includegraphics[width=0.860\textwidth]{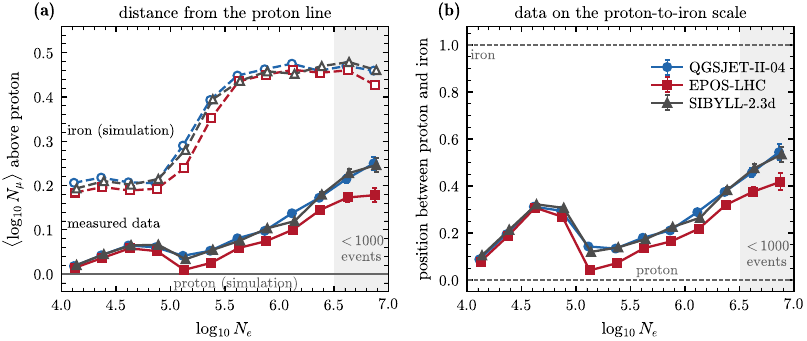}
\caption{At fixed electron size the measured mean muon size lies inside every model's proton-to-iron bracket. (a) Mean $\log_{10}N_\mu$ above each model's proton value, for the data (solid) and iron (dashed). (b) The data on each model's proton-to-iron scale ($0$ proton, $1$ iron). Zenith angle below $30^\circ$; error bars are the standard errors of the data only; grey: bins with fewer than $1{,}000$ events. The dip near $\log_{10}\Ne\approx5.1$--$5.4$ should not be read as composition: the bracket widens and the data's offset from proton changes there, for reasons not established (see text)}
\label{fig:measured}
\end{figure*}

\paragraph{Data.} The measured release contains $\epMeasuredTotal$ events with
$14\le\log_{10}(E_{\rm rec}/\mathrm{eV})\le18$; we use a random $\epMeasuredKeep\%$ of them. At the lowest
reconstructed energies ($14\le\log_{10}(E_{\rm rec}/\mathrm{eV})<14.2$) only $\epThrData\%$ of the measured
events have $\log_{10}\Ne<3.5$, against $\epThrProtonMin$ to $\epThrProtonMax\%$ for pure protons in
the reweighted simulation, and more for heavier primaries. No composition produces so few small
showers, so the measured release carries a selection in shower size that the simulation release
does not reproduce, consistent with the trigger: the array reaches full detection efficiency only
for $\log_{10}\Ne>4$ \citep{kascadeaniso}. We therefore apply $\log_{10}\Ne\ge\epNeCut$, the
full-efficiency threshold, identically to data and simulation, which leaves $\epMeasuredN$ measured
events. The common cuts are thus $14\le\lg(E_{\rm rec}/\mathrm{eV})<18$, zenith angle below $40^\circ$,
core within $91$\,m of the array centre, $\log_{10}\Ne\ge\epNeCut$ and finite reconstructed quantities;
the comparison at fixed electron size below adds zenith below $30^\circ$. Because the cut is in electron size it removes
more heavy than light showers near the lowest reconstructed energies, in data and simulation alike. The closure test above was rerun with the
same cut.

\paragraph{Data against simulation.} In every decade and for every model the data's mean
$\log_{10}N_\mu$ lies between those of pure proton and pure iron (table~\ref{tab:measured};
appendix~\ref{app:measured}). In the decades
$10^{14}$--$10^{15}$, $10^{15}$--$10^{16}$ and $10^{16}$--$10^{17}$\,eV the data's muon-size distribution is
closest to that of \epMeasClosestOne, \epMeasClosestTwo{} and \epMeasClosestThree{} on all three models. The distances are an order of magnitude larger than in the
closure test, as expected when the true composition is not this mixture. For the other
quantities the Wasserstein-1 distance between data and the mixture is $\epMeasWNeMin$ to
$\epMeasWNeMax$ in $\log_{10}\Ne$, $\epMeasWAgeMin$ to $\epMeasWAgeMax$ in age and $\epMeasWZeMin^\circ$ to
$\epMeasWZeMax^\circ$ in zenith angle, over all decades and models (values per decade and composition
are in the results files). At fixed electron
size (figure~\ref{fig:measured}) the data sit between $\epMeasZLowMin$ and $\epMeasZLowMax$ of the
way from proton to iron for $\log_{10}\Ne$ up to $5.5$, against $\epMeasZMixMin$ to
$\epMeasZMixMax$ for the mixture, and the value rises with size above $\log_{10}\Ne\approx5.25$,
to $\epMeasZHiMin$ to $\epMeasZHiMax$ in the highest bin ($\epMeasZHiN$ events). Every bin lies inside
the bracket for every model. This is a consistency check of the simulations on which the audit
rests, not a composition measurement; the light composition below the knee and the increase
above it that it suggests agree qualitatively with KASCADE's unfolded results \citep{kascade2005}. The simulated proton-to-iron interval itself widens from about $0.2$ to about $0.45$ in $\log_{10}N_\mu$ between $\log_{10}\Ne\approx5$ and $5.75$, and the offset of the data from proton also changes there, for a reason we have not established; the structure of the scale near that size, including the dip near $\log_{10}\Ne\approx5.1$--$5.4$ in figure~\ref{fig:measured}, should not be read as composition.

\paragraph{The probe on measured data.} Gradient boosting is trained on each simulation (after the
common cuts, over the whole energy range, unweighted) on $\phi=(\log_{10}\Ne,s,X_c,Y_c)$ together with
$M=(\log_{10}N_\mu,\theta)$, and the probe increment of $M$ is computed on its outputs for the measured
events (five random folds) and for held-out simulation showers resampled to the knee spectrum. For a
fixed trained function, agreement means that data and simulation have the same joint structure of
$(\phi,M)$ as seen through that function; it is not a test of the energy scale. The estimator scores a probe increment of $\epMeasIncDataMin$ to $\epMeasIncDataMax$ on
the measured events against $\epMeasIncSimMin$ to $\epMeasIncSimMax$ on held-out simulation, a
difference of at most $\epMeasIncDiffMax$, within the scale of the closure test. A
simulation-trained estimator uses the muon channel on measured showers as it does on simulated
ones, so the structure that the audit found in simulation is at least not contradicted by the
data in the one respect that can be checked without labels.

\section{GRAPES-3: a summary with nothing beyond it}
\label{sec:grapes}

\begin{figure*}[tb]
\centering
\includegraphics[width=0.700\textwidth]{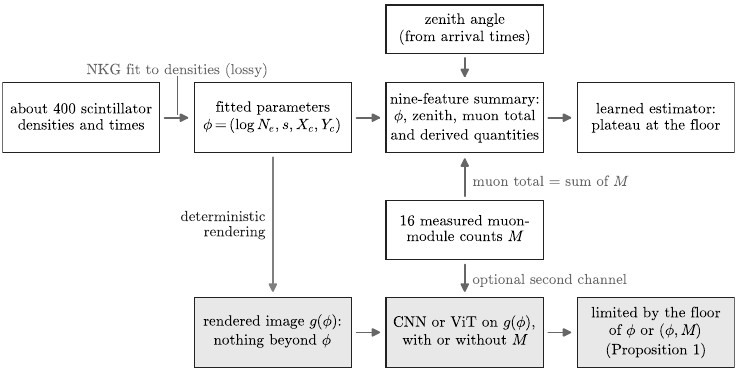}
\caption{The GRAPES-3 situation. The standard reconstruction replaces about 400 scintillator signals by four fitted parameters $\phi$; the nine-feature summary is built from $\phi$, the zenith angle and the muon total. An image rendered from $\phi$ is a deterministic function of $\phi$, so its squared-loss Bayes risk cannot be lower (Proposition~\ref{prop:bound}); only inputs outside $\phi$, such as the $16$ muon-module counts $M$ or the zenith angle, can add information}
\label{fig:pipeline}
\end{figure*}

\begin{table*}[tb]
\centering\small
\caption{The GRAPES-3 rendering experiment (60{,}000 training events). Shower cross-fit: every event is
predicted by a model that saw no copy of its shower (primary); random split: the archived protocol, which
leaks shower copies. Rows C$'$ and C$''$ are tabular controls added afterwards; the predicted column was written down before
rows A to D were run. The four-parameter floor estimate is
$0.1219$. Under the cross-fit the CNN on the rendered image is $0.0003$ below gradient boosting on the four
parameters (paired interval $[0.0001,0.0005]$): a difference between learners on the same information, which
Proposition~\ref{prop:bound} permits}
\label{tab:render}
\begin{tabular}{llccc}
\toprule
 & input and model & predicted & random split & shower cross-fit \\
\midrule
A & four fit parameters $\phi$, gradient boosting & $0.1234$ & $0.1230$ & $0.1230$ \\
B & CNN on image rendered from $\phi$ & $0.1234$ & $0.1226$ & $0.1227$ \\
C & CNN on rendered image $+$ 16 measured muon counts & $0.1063$ & $0.1063$ & $0.1064$ \\
C$'$ & $\phi$ $+$ 16 muon counts, gradient boosting & --- & $0.1071$ & --- \\
C$''$ & as C$'$ $+$ zenith angle & --- & $0.1029$ & --- \\
D & nine features, gradient boosting & $0.1019$ & $0.1022$ & $0.1024$ \\
\bottomrule
\end{tabular}
\end{table*}

\noindent\textbf{Key result.} Networks on images rendered from the fit do no better than the fit; only a measured channel helps.

GRAPES-3 shows the case in which the summary is all there is: the exported events contain the fit
parameters, the zenith angle and the muon-module counts, but no station signals. Here the audit can
illustrate its exact guarantee directly. Estimators of different inductive
bias agree, and cross-fitted gradient boosting ($\cfGB$) is between $\cfExcessAllMin\%$ and
$\cfExcessAllMax\%$ above the estimated floor of the nine features over every estimator setting
($\floorTest$ for the reference setting, shower-level interval $[\floorTestLo,\floorTestHi]$).
The ladder has the KASCADE shape: the zenith angle and the total muon count lower $\sigma$ (by
$\cfLadZen$ and $\cfLadMu$), and the pattern of the $16$ muon-module counts across the telescope
adds nothing resolvable to their total (paired interval $[\cfLadModLo,\cfLadModHi]$; permuting the
counts within each event makes no difference, $[\cfPermLo,\cfPermHi]$).

\paragraph{A rendering of the fit is not a measurement.} Figure~\ref{fig:pipeline} shows why an
image built from the fit cannot add information. An earlier network on this sample, a
$3.2$-million-parameter hybrid of convolutional and transformer stages given an image of
``particle density, arrival time and muon multiplicity'', reached $\sigma=\sigmaCNNViT$, level with
ridge regression. The exported files contain no per-scintillator densities or times, so a
density image built from them can only be the NKG function evaluated at detector positions, a
deterministic function of the fit (figure~\ref{fig:gallery}).

We reconstructed that input and wrote down the outcomes in
advance (table~\ref{tab:render}). Proposition~\ref{prop:bound} says that the rendered image cannot
lower the population Bayes risk below that of the four parameters it was made from; it says nothing
about how well two particular finite learners use that information. Empirically, under the leak-free
cross-fit by shower, the CNN on the rendered image reaches $\cfRendB$ against $\cfRendA$ for gradient
boosting on the four parameters, both within $1\%$ of the four-parameter floor estimate $\bayesFour$; the
CNN is $\cfRendBA$ lower (interval $[\cfRendBALo,\cfRendBAHi]$), a difference between learners on the same
information, not new information. The same holds in each of $\seedN$ seeds of the random-split
protocol (range $\seedBmin$ to $\seedBmax$), and a vision transformer on the image is level with the
CNN; linear and fully connected models on the pixels, which invert the rendering less well, do worse.
Adding the measured muon counts as a second channel gives $\cfRendC$; the remaining gap to the nine
features ($\cfRendD$) is the arrival direction, which the image does not contain.

On KASCADE, where the
station image is a detector observable rather than a rendering of the fit, the same kind of network improves on the summary
(section~\ref{sec:stations}); on GRAPES-3, where the image is rendered, it does not. This is a
corroborating illustration of Proposition~\ref{prop:bound} on one size slice, not an independent
validation of the whole audit. The degeneracy probe, run on the earlier network's archived
predictions before the reconstruction, had pointed to this: its output scores
$\cfsProbeArchPhi$ on the fit parameters and $\cfsProbeArchPhiM$ with the muon counts and zenith
angle, the signature of a function of those inputs (figure~\ref{fig:probemap}).

\paragraph{An inherited selection.} GRAPES-3 analyses commonly accept events with shower age
$0.90\le s\le1.15$, a cut designed for full-range reconstruction. Within this narrow size slice it
keeps $\epSelAcc\%$ of the events. Three comparisons answer different questions, and we keep them
apart (shower cross-fit throughout). Operationally, an analysis that applies the cut ends with
$\sigma=\epSelAccS$ on the events it keeps, against $\epSelAll$ without the cut, and with a spread of
the mean bias across primaries of $\epSelSpreadAccAll$ against $\epSelSpreadAll$; but these are
different populations. On the fixed population of accepted events, training within the cut instead of
on all events changes $\sigma$ by $\epSelDiff$ (interval $[\epSelDiffLo,\epSelDiffHi]$), as it must:
the cut is a function of the summary, so restricting training to it cannot lower the Bayes risk on
the events it keeps. What the cut does is select the harder half. The floor estimate of the nine
features is $\epSelFloorAcc$ on the accepted events ($[\epSelFloorAccLo,\epSelFloorAccHi]$) and
$\epSelFloorRej$ on the rejected ones ($[\epSelFloorRejLo,\epSelFloorRejHi]$), where gradient boosting
reaches $\epSelRej$. The cut therefore does not damage the reconstruction; it discards the events the
summary allows to be reconstructed best, which is counterproductive for an analysis that wants the
best resolution in this slice. On KASCADE an analogous central-age selection likewise leaves the
resolution on its own events unchanged by retraining (by at most $\epKSelRetrainMax\%$), while the
population it keeps differs in $\sigma$ by $\kcdcDeltaMin\%$ to $\kcdcDeltaMax\%$ with a sign that
depends on the size slice. The audit therefore states which estimand a selection is judged by, and
re-measures it in the regime where it is applied.

\section{Discussion}
\label{sec:discussion}

\paragraph{Where the error comes from.} For KASCADE-like arrays the limit of summary-based energy
estimation is not the learner. Standard learners on the nine features are at the estimated floor, to within that estimate's precision of about two per cent below $10^{16}$\,eV, and within its wider precision up to $10^{17}$\,eV, on all three interaction models. What
remains has two sources whose balance shifts with energy. At low energy most of the error goes with
the reconstructed particle numbers. At higher energy most of it survives even with true sizes, which
points to shower-to-shower fluctuations and the unknown primary mass. These statements are for a known
energy decade; without it the same pattern holds, but an estimator in use has the larger resolutions
of table~\ref{tab:fullrange}, and training on the generated spectrum without weighting fails between $10^{16}$ and $10^{17}$\,eV (section~\ref{sec:fullrange}).

\paragraph{What the station signals change.} The station signals carry information the summary lacks,
so the summary is not sufficient for the energy. Yet four encoders recover only $\btMeanMin$ to
$\btMeanMax\%$ of what reconstruction loses, on every model between $10^{14}$ and $10^{16}$\,eV, and the one
trained seed of a larger pre-registered run does not do much better. A large gain from networks of
this kind therefore looks unlikely, but whether a different model of the station signals would recover
much more is open. The oracle shows
what the reconstructed totals cost, not how much of that is reachable, since part of the loss may come
from the sparse sampling of the shower by the array, which no function of the station signals undoes.

\paragraph{Interaction models and a second reconstruction.} The three models differ in muon content and shower development, and
their absolute resolutions differ slightly (table~\ref{tab:decades}). The structure of the result
does not: the plateau at the estimated floor (within its precision), the ladder, the energy dependence of the loss and the
small recovery by station signals hold for all three. The conclusions are therefore not artefacts of
one model, while any composition analysis built on these estimators inherits the usual model
dependence.
The KASCADE-Grande combined reconstruction, on a larger array, repeats the plateau, the falling
reconstruction share and the gain over a fixed formula on all three models (section~\ref{sec:comb}); its
one failed check comes from the smallest neighbour range in a sparse decade. Because it reuses the KASCADE
showers, it argues against an artefact of one array or one reconstruction, but not of the simulation.

\paragraph{What an array collaboration can do.} Three practices follow from the results.
\begin{itemize}
\item \emph{Estimate the floor before enlarging a model.} On KASCADE below $10^{16}$\,eV, with the energy decade
known, gradient boosting was within $\epSensLowAbsMax\%$ of the estimated floor of the summary
(section~\ref{sec:plateau}); a larger network on the same inputs is not expected to gain more than
about two per cent there, while the station signals still held
unused information
(section~\ref{sec:stations}).
\item \emph{Record which inputs were measured and which were computed.} On GRAPES-3 an image rendered
from the fit was treated as a richer input, and it could not be one (section~\ref{sec:grapes}). A
channel-by-channel record would have shown this before any training.
\item \emph{Re-measure inherited selections where they are applied.} The GRAPES-3 age cut, designed
for full-range reconstruction, discards the events that the summary reconstructs best within the
slice studied here (section~\ref{sec:grapes}).
\end{itemize}

\paragraph{Beyond these arrays.} Proposition~\ref{prop:bound} applies wherever fitted parameters are
distributed and image-like inputs are rebuilt from them. The floor estimate, the ladder and the probe
transfer as procedures whose settings must be re-established for each data set; no number transfers.
The audit is also cheap: on a laptop CPU the floor estimates, ladder and probe for one testbed take
minutes, and the only network trainings are those of the inputs being audited.

\paragraph{Limitations.} Several limits qualify these conclusions.
\begin{itemize}
\item \emph{The floor is an estimate, not a bound.} Its intervals under-cover, a calibration of them
failed on held-out noise levels, and its accuracy was measured on a target that imitates the real one,
not on the real one. Between $10^{16}$ and $10^{17}$\,eV its settings move the comparison with
boosting by $\epSensThirdLo$ to $\epSensThirdHi\%$. On the known-truth benchmark it errs by up to
$\synTMVerr\%$ in eight dimensions. In the data-limited top decade it even falls below the oracle
(table~\ref{tab:decades}), which no true floor of the summary can do, so it is biased low where events
are sparse.
\item \emph{Squared loss and the energy only.} Channels that add nothing here may matter for
composition or for calibrated uncertainties.
\item \emph{Energy range.} The station-level analysis covers two decades ($\lgE\in[14,16]$), and the summary-level audit reaches $10^{17}$\,eV on two reconstructions. Above $10^{17}$\,eV the public KASCADE and KASCADE-Grande simulations hold too few showers, and the floor estimate falls below the oracle there (section~\ref{sec:comb}); a dedicated simulation would be needed.
\item \emph{The oracle is not observable.} It uses particle numbers that no detector measures.
\item \emph{One simulation domain.} Both testbeds are ground arrays simulated with CORSIKA, so the second array is independent in detector but not in domain, and its simulation shares QGSJET-II-04 with the main KASCADE set. The KASCADE-Grande sets reuse the KASCADE showers, model by model ($\cbOverlapMin$--$\cbOverlapMax\%$ shared). The synthetic benchmark is the only check outside that domain.
\item \emph{GRAPES-3 coverage.} The GRAPES-3 sample is one size slice. The archived network's code
and weights are lost, so our account of that network's input is a conjecture, supported by the
reconstruction and by the probe.
\end{itemize}

\section{Conclusion}
\label{sec:conclusion}

When an energy estimator for an air-shower array is trained on the fitted shower summary, the
summary limits what it can do, and the usual training diagnostics do not show that limit. We
separated the two with an audit built around a learner-independent estimate of the floor, the lowest
error the summary allows, and an exact statement about inputs computed from the fit.

What we learned is where the error lives. On KASCADE simulation with three interaction models,
within a known energy decade and below $10^{16}$\,eV, learned estimators on the summary sit within about two per cent of the estimated floor, so better learners on the same summary have little left to gain, to within the estimate's precision. The limit is what the summary records, and its
character changes with energy: the share of error variance tied to reconstruction falls from about
$\epOneShareMax\%$ below $10^{15}$\,eV to about $\epThreeShareMax\%$ between $10^{16}$ and $10^{17}$\,eV,
in the same way for every model. On the KASCADE-Grande combined reconstruction, which starts at $10^{15}$\,eV, it falls in the same direction on all three models, from about $\cbTwoShareMax\%$ to about $\cbThreeShareMax\%$ or less. The collaboration's fixed-formula estimate is $\epKcdcExOneMin$ to $\epKcdcExThreeMax\%$ above the estimated floor within a known decade, increasingly so with energy, and a standard learner trained over the whole range still improves on it by $\epFrKcdcOverMin\%$ or more. The station signals carry information beyond the summary, but four encoders on three models close only about a tenth of the gap to the oracle ($\btMeanMin$--$\btMeanMax\%$), and on GRAPES-3 inputs rendered from the fit do no better than the fit, as the guarantee leads one
to expect.

Some things are not established. The floor estimate's accuracy is known only on a semi-synthetic
target, and its intervals are lower limits. Apart from the label-free checks on measured KASCADE data, all results are on CORSIKA simulation, so the agreement across detectors and interaction models is not a validation outside that domain. The larger station run covers one model and one energy
window at a $20$-epoch budget, two of its three seeds and its control collapsed to boosting, and so it
rests on one seed; whether a different model
of the station signals could close much more of the oracle gap remains open.

Two directions follow. The first is a dedicated simulation above $10^{17}$\,eV, where the public releases hold too few showers to test whether the reconstruction share keeps falling with energy. The second is practical and applies to any array: before enlarging a model
or redesigning its input, estimate what the representation allows, and record which inputs were
measured.

\FloatBarrier
\backmatter

\bmhead{Acknowledgements}
We thank Dr.~Arun Nayak of the Institute of Physics, Bhubaneswar, for providing the simulated GRAPES-3
events used in this work, and the KASCADE Cosmic-ray Data Centre for its public simulation and data releases.

\section*{Statements and Declarations}

\bmhead{Funding} This research received no specific grant from any funding agency.

\bmhead{Competing interests} The authors have no competing interests to declare that are relevant
to the content of this article.

\bmhead{Data availability} The KASCADE simulations (QGSJET-II-04, EPOS-LHC and SIBYLL-2.3d sets, KASCADE component), the KASCADE-Grande COMBINED simulations for the same three models, and the measured data are public through the KASCADE Cosmic-ray Data Centre
(\url{https://kcdc.iap.kit.edu}) \citep{kcdc,kcdcsim}; the station positions are tabulated in its
simulation manual. The two GRAPES-3 simulation files ($\Nroot$ events) were made available to the authors
for research use and may not be redistributed; requests for access should be directed to the
GRAPES-3 Collaboration.

\bmhead{Code availability} The analysis code, the pipeline that regenerates every number, table
and figure from the data, and the script that checks every number in the text against the results
files are at \url{https://github.com/SamarjeetMalik/summary-representation-limits}; the version used
for this manuscript is tagged \texttt{epjc-submission-v13}. The README describes how to regenerate every
table and number macro from the stored results files without the data.

\bmhead{Author contributions} Both authors conceived and designed the study. Mugdha Raje Ashiya
carried out the GRAPES-3 analysis. Samarjeet Malik carried out the KASCADE analysis, wrote the
analysis software and drafted the manuscript. Both authors revised the manuscript critically and
read and approved the final version.

\onecolumn
\newcounter{mainfigures}\setcounter{mainfigures}{\value{figure}}
\edef\maintablecount{\arabic{table}}
\begin{appendices}
\renewcommand{\thetable}{\arabic{table}}
\setcounter{table}{\maintablecount}
\gdef\thefigure{\arabic{figure}}
\setcounter{figure}{\value{mainfigures}}

\section{Proofs}
\label{app:proofs}

For a sub-$\sigma$-algebra $\mathcal G$ the squared-loss Bayes predictor is $\E[y\mid\mathcal G]$,
and for nested $\mathcal G\subseteq\mathcal H$ the tower property gives
\begin{equation}
\begin{split}
\E\bigl[(y-\E[y\mid\mathcal G])^2\bigr]&=\E\bigl[(y-\E[y\mid\mathcal H])^2\bigr]\\
&\quad+\E\bigl[(\E[y\mid\mathcal H]-\E[y\mid\mathcal G])^2\bigr],
\end{split}
\label{eq:pythag}
\end{equation}
because the cross term vanishes on conditioning on $\mathcal H$.

\emph{Proposition~\ref{prop:bound}.} $\sigma(g(T))\subseteq\sigma(T)$; apply \eqref{eq:pythag} with
$\mathcal G=\sigma(g(T))$ and $\mathcal H=\sigma(T)$. The infima are attained by the conditional
expectations, so their difference is the second term, which is non-negative and vanishes exactly
when $\E[y\mid g(T)]=\E[y\mid T]$ almost surely.

\emph{Proposition~\ref{prop:channel}.} The equality is \eqref{eq:pythag} with $\mathcal
G=\sigma(T)$, $\mathcal H=\sigma(T,M)$. For fixed $(T,M)$ let $P$ and $Q$ be the conditional
distributions of $y$ given $(T,M)$ and given $T$. With $c=(a+b)/2$,
\begin{align*}
|\E_P y-\E_Q y|&=\Bigl|\int(y-c)\,\mathrm d(P-Q)\Bigr|\\
&\le(b-a)\,\mathrm{TV}(P,Q)\\
&\le(b-a)\sqrt{\mathrm{KL}(P\|Q)/2}
\end{align*}
by Pinsker's inequality \citep{coverthomas}, with
$\mathrm{TV}(P,Q)=\sup_A|P(A)-Q(A)|=\tfrac12\|P-Q\|_1$ and KL in nats; squaring and averaging over $(T,M)$, with
$\E[\mathrm{KL}(P\|Q)]=I(y;M\mid T)$, gives the bound.

\section{GRAPES-3 protocol}
\label{app:grapes}
\label{sec:protocol}

Each CORSIKA shower is thrown at several core positions, so a random split of events leaks
copies of test showers into training. Dropping training events whose shower appears in the test
set removes the leak but biases the training sample towards showers thrown few times. We therefore
cross-fit by shower: showers are assigned to five folds and each fold's events are predicted by a
model trained on the other four. The leak under a random split is at most $\cfLeakMax$ in $\sigma$,
and no conclusion depends on the protocol. Table~\ref{tab:floor} lists the estimators on the nine
features.

\begin{figure}[H]
\centering
\includegraphics[width=0.791\textwidth]{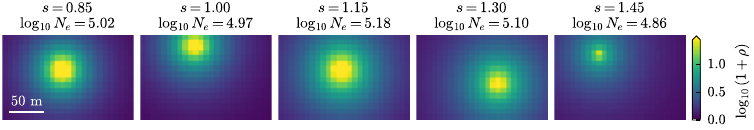}
\caption{Images rendered from the fit look like detector images but contain nothing beyond the four fitted parameters. NKG densities on a rectangular $25\times16$ grid with $8$\,m pitch, for five events of increasing shower age $s$; the events also differ in size (printed) and core position. $\rho$ is the NKG density in m$^{-2}$; the colour scale is capped at the 99th percentile so that the shower halo is visible}
\label{fig:gallery}
\end{figure}

\par\medskip\noindent\begin{minipage}{\linewidth}
\centering\footnotesize
\captionof{table}{Estimators and floor estimates on the nine-feature summary, all scored on the same 34{,}356
GRAPES-3 evaluation events. $\sigma$ is the standard deviation of $\hat y-y$ with $y=\lgE$. For
the estimators the two columns are the two training protocols of section~\ref{sec:protocol}: the
random complement (mean of five seeds), and a five-fold cross-fit by shower in which no model
sees a copy of a shower it is scored on. The lower block gives Gamma-test estimates
$\hat\sigma_{\mathrm{B}}$ of the square root of the squared-loss Bayes risk of the same features,
with $95\%$ intervals; they train no regressor and are estimates, not bounds. The intervals
resample showers, except in the last row.}
\label{tab:floor}
\begin{tabular}{lcc}
\toprule
 & random split & shower cross-fit \\
\midrule
\multicolumn{3}{l}{\emph{Learned estimators, nine features}} \\
Ridge (10 coefficients) & $0.1066$ & $0.1067$ \\
Random forest & $0.1034$ & $0.1036$ \\
Gradient boosting & $0.1019$ & $0.1021$ \\
MLP $64{\times}64$ & $0.1022$ & $0.1031$ \\
XGBoost & $0.1024$ & --- \\
$k$-nearest neighbours ($k=50$) & $0.1041$ & --- \\
\midrule
\multicolumn{3}{l}{\emph{References}} \\
Size--age calibration (2 coefficients) & $0.1284$ & --- \\
Archived CNN/ViT, rendered input (corroborative) & $0.1063$ & --- \\
\midrule
\emph{Estimated floor $\hat\sigma_{\mathrm{B}}$, nine features (Gamma test)} & estimate & interval \\
evaluation events as queries, neighbours from all other showers & $0.1018$ & $[0.1006,0.1030]$ \\
evaluation events only & $0.1007$ & $[0.0995,0.1018]$ \\
evaluation queries, six non-redundant coordinates & $0.1015$ & $[0.1003,0.1027]$ \\
all 171{,}776 events (event-level interval) & $0.1003$ & $[0.0998,0.1009]$ \\
\bottomrule
\end{tabular}
\end{minipage}\par\medskip

\begin{figure}[H]
\centering
\includegraphics[width=0.868\textwidth]{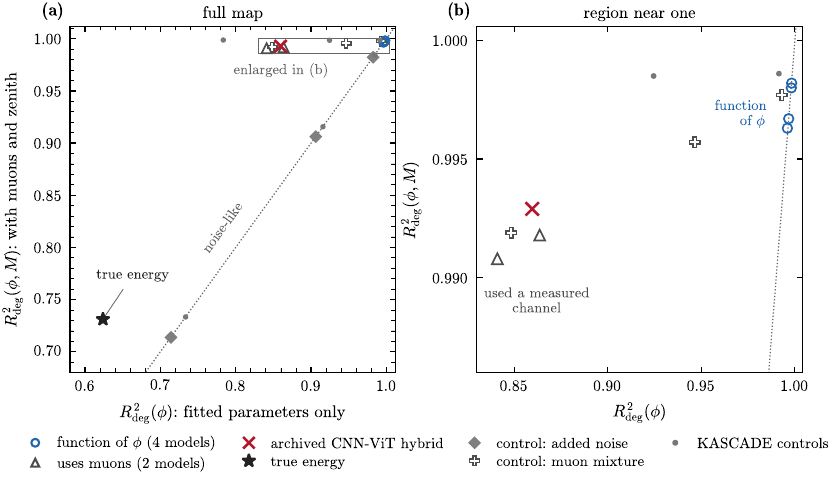}
\caption{The degeneracy probe. Each point other than the small grey dots is a GRAPES-3 output or reference (``archived'' is the earlier CNN-ViT hybrid's stored predictions) probed on the fitted parameters alone (horizontal) and with the muon counts and zenith angle (vertical): (a) full map, (b) the region near one, enlarged. Functions of $\phi$ sit at the top right, but a small admixture can sit beside them (a $10\%$ muon mixture scores $0.993$ on $\phi$ alone), so a high score is necessary, not sufficient; added noise moves an output down the diagonal; information from outside $\phi$ moves it left at the top. The true energy is not a function of these inputs, so it scores below one on both axes. Grey dots are the corresponding controls on KASCADE, whose muon mixture also uses $\theta$; the isolated one at $(0.78,1.00)$ is its $50\%$ mixture. Here $M$ is the 16 muon counts with the zenith angle on GRAPES-3, and $(\log_{10}N_\mu,\theta)$ on KASCADE}
\label{fig:probemap}
\end{figure}

\section{Known-truth benchmark}
\label{app:synth}

\begin{figure}[H]
\centering
\includegraphics[width=0.838\textwidth]{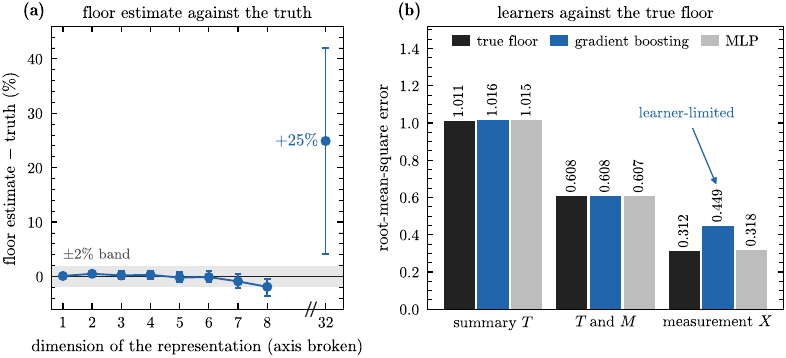}
\caption{The known-truth benchmark. (a) Error of the Gamma-test floor estimate against the true floor, with its stored interval, as the representation grows from one coordinate of the summary to the summary with two further channels (eight dimensions) and to the raw measurement ($32$, axis broken). The $\pm2\%$ band is a visual guide; the interval at eight dimensions lies below zero and the one at $32$ dimensions above it ($+4\%$ to $+42\%$). (b) Gradient boosting and an MLP against the true floor; on the raw measurement boosting is learner-limited. Values are on a finite test sample, so a learner can print slightly below the true floor}
\label{fig:benchmark}
\end{figure}

A measurement $X\in\mathbb R^{32}$ is Gaussian with covariance $\Sigma=Q\Lambda Q^\top$ ($Q$ a random
orthogonal matrix, $\Lambda_{jj}=j^{-1/2}$), the summary is a fixed lossy linear map $T=AX\in\mathbb R^{6}$
(entries of $A$ drawn from $N(0,1/32)$; nested subsets are its first $k$ coordinates), and $y=h(w^\top X)+\tau(0.5+V)\xi$ with
$h(u)=u+0.5\sin2u$, $w$ normalised so that $\operatorname{Var}(w^\top X)=1$, $\tau=0.30$, $\xi\sim N(0,1)$; a
further measured channel $M=m^\top X+0.30\,\eta$ with $m=w+0.8\,\zeta/\sqrt{32}$ ($\eta$, $\zeta$ standard
normal), which carries information about $w^\top X$; and a noise-scale channel $V\sim U(0,1)$. There
are $170{,}000$ samples, the last $34{,}000$ held out. The rendered image for the CNN row is a Gaussian
blob on a $16\times25$ lattice whose amplitude, width and centre are smooth functions of four
standardised coordinates of $T$, plus a linear gradient set by the other two; no measurement enters
it. The Gamma-test row is omitted for the image, whose dimension is far beyond the estimator's
range; its true floor equals that of $T$ by Proposition~\ref{prop:bound}. Every Bayes risk reduces to a Gauss--Hermite integral. The Gamma-test
estimate is within $\synLowDimErr\%$ of the truth on the summary and its nested subsets (one to six
dimensions), $\synTMerr\%$ low with $M$ added and $\synTMVerr\%$ low with $V$ in eight dimensions,
and $\synXerr\%$ too high on the $32$-dimensional measurement (table~\ref{tab:synth},
figure~\ref{fig:benchmark}). Gradient
boosting on $T$ is $\synGBexcess\%$ above the true floor. On the measurement an MLP reaches
$\synMlpX$ against a floor of $\synFloorX$ while gradient boosting stops at $\synGBX$: a
learner-limited case, which the audit is built to tell apart and which only the known truth
resolves here. The six-dimensional summary matches the six non-redundant KASCADE features in
dimension, so the benchmark validates the estimator in the relevant low-dimensional regime, not in
general; its accuracy in the actual KASCADE geometry is measured separately
(table~\ref{tab:floorbench}).

\par\medskip\noindent\begin{minipage}{\linewidth}
\centering\footnotesize
\captionof{table}{The audit on a benchmark whose Bayes risks are known. For each representation: the true
floor $\sqrt{\Rstar}$, the Gamma-test estimate with its relative error, and the root-mean-square
error of the best learned estimator (gradient boosting or an MLP). The last row is a CNN on an
image computed from $T$; its floor is that of $T$.}
\label{tab:synth}
\begin{tabular}{lcccc}
\toprule
Representation & true floor & $\hat\sigma_{\mathrm{B}}$ & error & best learned \\
\midrule
summary $T$ (six numbers) & $1.011$ & $1.010$ & $-0.1\%$ & $1.015$ \\
$T$ and measured channel $M$ & $0.608$ & $0.603$ & $-0.9\%$ & $0.607$ \\
$T$, $M$ and noise-scale channel $V$ & $0.608$ & $0.596$ & $-1.9\%$ & $0.608$ \\
measurement $X$ (32 numbers) & $0.312$ & $0.390$ & $+24.9\%$ & $0.318$ \\
\midrule
image rendered from $T$, CNN & $1.011$ & --- & --- & $1.029$ \\
\bottomrule
\end{tabular}
\end{minipage}\par\medskip

\section{Accuracy of the floor estimate, and the ladder by decade}
\label{app:tables}

This appendix gives the detail behind the short account in section~\ref{sec:plateau}.

\begin{figure}[H]
\centering
\includegraphics[width=0.884\textwidth]{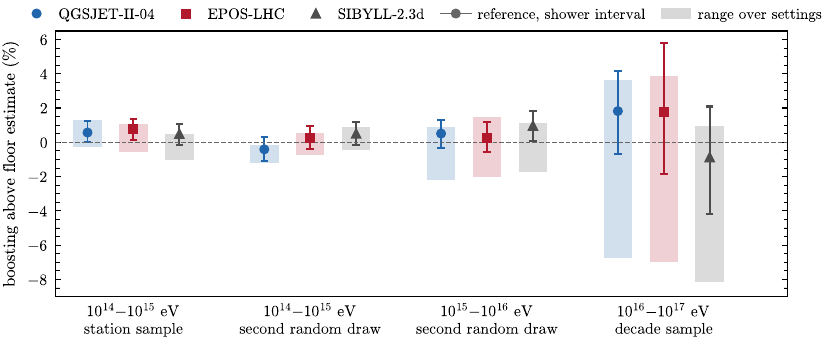}
\caption{Below $10^{16}$\,eV gradient boosting stays within about two per cent of the floor estimate under every estimator setting; between $10^{16}$ and $10^{17}$\,eV the spread is wider. Point and error bar: the reference setting with a joint shower interval, a lower limit on the uncertainty because it under-covers in the benchmark (table~\ref{tab:floorbench}); shaded bar: the range over all settings (table~\ref{tab:floorsens}). The station sample and the second random draw are described in the text; the decade sample is that of table~\ref{tab:decades}, with intervals from a separate bootstrap}
\label{fig:floorsens}
\end{figure}

\paragraph{Sensitivity to settings.} The estimate varies for two reasons: which
showers happen to be sampled, and which estimator settings are used (neighbour range, feature
scaling, metric, coordinates). Table~\ref{tab:floorsens} gives both for
every model, for each decade and for the station-level sample, and figure~\ref{fig:floorsens} shows
them. The decade rows use a second random draw from the release, which overlaps the sample of
table~\ref{tab:decades} and is the same set of events between $10^{16}$ and $10^{17}$\,eV.

In figure~\ref{fig:floorsens} the station sample is the station-level sample of about
$410{,}000$ events; the sensitivity run's floors on this sample differ from those of the station analysis (table~\ref{tab:stations}) by $0.0002$ to $0.0004$ across
models ($0.0002$ for the QGSJET-II-04 sample of figure~\ref{fig:kcdcaudit}). The intervals in figure~\ref{fig:floorsens} and those
in figure~\ref{fig:decades} come from separate joint bootstrap runs ($500$ and $300$ replicates), so
they differ slightly for the same events.

The settings matter more than the sampling. In $\epSensSetWider$ of the $\epSensN$ cases they move
the estimate by more than its shower interval, and the Manhattan metric alone moves it by up to
$\epSensManhattanMax\%$.

The comparison with gradient boosting survives this variation (table~\ref{tab:floorsensgb}). At the reference setting boosting is
within $\epSensRefAbsMax\%$ of the estimate in every case; its interval excludes zero in
$\epSensCiExcl$ of the $\epSensN$, by at most $\epSensCiExclMax\%$ at the lower end. Over all
settings, boosting is within $\epSensLowAbsMax\%$ of the estimate below $10^{16}$\,eV, sometimes
above it and sometimes below; there the settings range spans zero in all but one of the nine
cases. Between $10^{16}$ and $10^{17}$\,eV, where fewer events widen the
spread, boosting is between $\epSensThirdLo\%$ and $\epSensThirdHi\%$ from the estimate. The wide end there comes
almost entirely from the rank scaling; without it the range is $\epSensThirdNoRankLo$ to
$\epSensThirdNoRankHi\%$.

Two further checks concern the geometry of the features. Using the six non-redundant coordinates
instead of all nine moves the estimate by at most $\epSensSixMaxPct\%$, so the three derived features
do not distort the neighbour geometry. The direct nearest-neighbour value $\sqrt{\gamma(1)}$, without
the extrapolation in $\delta$, is $\epSensNnMin$ to $\epSensNnMax\%$ higher, most in the sparse decade,
as expected of an estimate that includes the change of the regression function between neighbours.

\paragraph{Accuracy in the KASCADE geometry.} Sensitivity to settings says how much the estimate
moves, not how far it is from the truth. To measure that, we need a target whose floor is known but
whose features look exactly like the real ones. We therefore keep the real nine features of each
sample, with its sample size and its shower copies, and replace the target by $g(X)+s(X)\,\xi$. Here
$g$ and $s$ are learned from the real target (gradient boosting for the conditional mean and for the
squared residual), so the regression surface and the heteroscedasticity are realistic, and $\xi$ is
standard normal, drawn once per CORSIKA shower. The true floor $\sqrt{\langle s^2\rangle}$ is then
known, and every setting can be scored against it over $\epBenchReps$ noise replicates
(table~\ref{tab:floorbench}).

The estimate is accurate but slightly high. Below $10^{16}$\,eV the reference setting reads high by
$\epBenchRefLowMin$ to $\epBenchRefLowMax\%$, and every setting except the rank scaling and the direct
1-NN value lies within $\epBenchLowAllMin$ to $\epBenchLowAllMax\%$ of the truth. In the sparse decade
$10^{16}$--$10^{17}$\,eV the reference reads high by $\epBenchRefThirdMin$ to $\epBenchRefThirdMax\%$
and the rank scaling by up to $\epBenchRankThirdMax\%$, which is where the wide settings range comes
from. Because the shower interval of the reference estimate ignores this small upward bias, it covers the true floor in only
$\epBenchCovMin$ to $\epBenchCovMax\%$ of the replicates instead of the nominal $95\%$. Correcting the
floor estimate for the measured bias puts boosting $\epBenchImpliedMin$ to $\epBenchImpliedMax\%$
above the corrected estimate in each sample below $10^{16}$\,eV.

\paragraph{Can the interval be calibrated?} A natural repair is to correct the estimate for the bias
measured on this target and to widen its interval until it covers $95\%$. We tested this repair with
a procedure and a pass mark fixed before the run (\texttt{PREREG\_v3.md} in the code repository). The
correction and the widening were fitted on the target above, pooled over the three interaction
models in each decade, and then checked on four noise families not used in the fit: the noise level
scaled by $0.6$ and by $1.5$, homoscedastic noise, and stronger heteroscedasticity, all on the same
regression surface. The pass mark was $92$--$97\%$ coverage. The calibrated interval reached only
$\calibCovMin$ to $\calibCovMax\%$, so all $\calibNfail$ decade groups failed.

The failure has a clear cause. At the lower noise level coverage fell to $\calibLowMin$ to
$\calibLowMax\%$, against $\calibRestMin$ to $\calibRestMax\%$ in the other three families. The
estimate's relative bias grows as the noise falls, so a correction fitted at one noise level does not
transfer to another. We therefore keep the uncalibrated shower interval and read it as a lower limit
on the uncertainty.

\paragraph{Tables.} Table~\ref{tab:ladderdec} gives the ladder by decade. Its last column uses the six non-redundant
features (fit, zenith angle and muon size), so it can differ slightly from the nine-feature
boosting column of table~\ref{tab:decades} where events are few.

\IfFileExists{tables/e_ladderdec.tex}{
\par\medskip\noindent\begin{minipage}{\linewidth}
\centering\small
\captionof{table}{The ladder by decade of true energy: $\sigma$ of gradient boosting on the four fit
parameters, with the zenith angle added, and with the muon size added}
\label{tab:ladderdec}
\footnotesize\setlength{\tabcolsep}{3.5pt}
\begin{tabular}{llccc}
\toprule
$\lg(E/\mathrm{eV})$ & model & fit & + zenith & + muon size \\
\midrule
$14$--$15$ & QGSJET-II-04 & $0.2017$ & $0.1753$ & $0.1481$ \\
 & EPOS-LHC & $0.2010$ & $0.1743$ & $0.1486$ \\
 & SIBYLL-2.3d & $0.2012$ & $0.1755$ & $0.1491$ \\
\midrule
$15$--$16$ & QGSJET-II-04 & $0.1579$ & $0.1149$ & $0.0870$ \\
 & EPOS-LHC & $0.1553$ & $0.1121$ & $0.0851$ \\
 & SIBYLL-2.3d & $0.1553$ & $0.1144$ & $0.0873$ \\
\midrule
$16$--$17$ & QGSJET-II-04 & $0.1213$ & $0.0808$ & $0.0615$ \\
 & EPOS-LHC & $0.1155$ & $0.0755$ & $0.0600$ \\
 & SIBYLL-2.3d & $0.1187$ & $0.0794$ & $0.0626$ \\
\midrule
$17$--$18$ & QGSJET-II-04 & $0.1022$ & $0.0720$ & $0.0573$ \\
 & EPOS-LHC & $0.1063$ & $0.0698$ & $0.0622$ \\
 & SIBYLL-2.3d & $0.1058$ & $0.0720$ & $0.0578$ \\
\bottomrule
\end{tabular}
\end{minipage}\par\medskip
}{}

\par\medskip\noindent\begin{minipage}{\linewidth}
\centering\footnotesize\setlength{\tabcolsep}{5pt}
\captionof{table}{Gradient boosting against the floor estimate on KASCADE simulation, for the
station-level sample and each decade of true energy: its resolution, the reference floor estimate,
its excess over that estimate in per cent with a joint shower interval, the range of the excess over
all estimator settings of table~\ref{tab:floorsens}, and the bias of the reference estimate measured
on a semi-synthetic target with a known floor in the same sample (table~\ref{tab:floorbench}). A
positive bias means the estimate reads high, so the excess of boosting over the true floor is larger
than over the estimate by about that amount. Shower intervals reflect sampling only and should be read
as lower limits on the uncertainty}
\label{tab:floorsensgb}
\begin{tabular}{llcccccc}
\toprule
 & & & & \multicolumn{3}{c}{boosting above floor est. (\%)} & bias of est. \\
$\lg(E/\mathrm{eV})$ & model & boosting & floor est. (ref.) & ref. & interval & all settings & (benchmark, \%) \\
\midrule
$14$--$15$ (stations) & QGSJET-II-04 & $0.1485$ & $0.1477$ & $+0.6$ & $[+0.0,+1.2]$ & $[-0.2,+1.3]$ & $+0.3$ \\
 & EPOS-LHC & $0.1484$ & $0.1473$ & $+0.8$ & $[+0.1,+1.4]$ & $[-0.6,+1.0]$ & $+0.3$ \\
 & SIBYLL-2.3d & $0.1495$ & $0.1488$ & $+0.5$ & $[-0.1,+1.1]$ & $[-1.1,+0.5]$ & $+0.2$ \\
\midrule
$14$--$15$ & QGSJET-II-04 & $0.1476$ & $0.1482$ & $-0.4$ & $[-1.1,+0.3]$ & $[-1.2,-0.2]$ & $+0.4$ \\
 & EPOS-LHC & $0.1483$ & $0.1480$ & $+0.2$ & $[-0.4,+0.9]$ & $[-0.7,+0.6]$ & $+0.1$ \\
 & SIBYLL-2.3d & $0.1500$ & $0.1492$ & $+0.5$ & $[-0.2,+1.2]$ & $[-0.4,+0.9]$ & $+0.4$ \\
\midrule
$15$--$16$ & QGSJET-II-04 & $0.0866$ & $0.0861$ & $+0.5$ & $[-0.3,+1.3]$ & $[-2.2,+0.9]$ & $+0.5$ \\
 & EPOS-LHC & $0.0853$ & $0.0851$ & $+0.2$ & $[-0.6,+1.2]$ & $[-2.0,+1.5]$ & $+0.6$ \\
 & SIBYLL-2.3d & $0.0875$ & $0.0866$ & $+1.0$ & $[+0.1,+1.9]$ & $[-1.7,+1.1]$ & $+0.5$ \\
\midrule
$16$--$17$ & QGSJET-II-04 & $0.0614$ & $0.0603$ & $+1.8$ & $[-0.7,+4.2]$ & $[-6.7,+3.7]$ & $+1.7$ \\
 & EPOS-LHC & $0.0596$ & $0.0586$ & $+1.7$ & $[-1.8,+5.8]$ & $[-7.0,+3.9]$ & $+1.7$ \\
 & SIBYLL-2.3d & $0.0623$ & $0.0628$ & $-0.9$ & $[-4.2,+2.1]$ & $[-8.1,+0.9]$ & $+2.4$ \\
\bottomrule
\end{tabular}
\end{minipage}\par\medskip

\par\medskip\noindent\begin{minipage}{\linewidth}
\centering\footnotesize\setlength{\tabcolsep}{2.6pt}
\captionof{table}{Sensitivity of the floor estimate on KASCADE simulation, for the station-level sample
($\lg(E/\mathrm{eV})\in[14,15]$) and for each decade of true energy ($\lg(E/\mathrm{eV})$, a uniform sample of at most $300{,}000$
events: a second random draw for the two lower decades, the events of table~\ref{tab:decades} for
$[16,17]$). Reference setting: test events as queries,
neighbours from other showers, standardised nine features, Euclidean metric, $K=10$; its interval
resamples showers. The next columns vary one setting at a time: the neighbour range $K$, three
feature scalings (standardised, robust, rank) at $K=10$, the Manhattan metric, and the six
non-redundant coordinates instead of nine; each passed the known-noise control. ``Direct 1-NN'' is
$\sqrt{\gamma(1)}$ without the extrapolation in $\delta$, biased high by the variation of the
regression function between neighbours}
\label{tab:floorsens}
\begin{tabular}{llcccccccc}
\toprule
$\lg(E/\mathrm{eV})$ & model & reference & shower interval & $K=5$--$50$ & 3 scalings & Manhattan & six coord. & direct 1-NN \\
\midrule
$14$--$15$ (stations) & QGSJET-II-04 & $0.1477$ & $[0.1467,0.1485]$ & $0.1473$--$0.1481$ & $0.1476$--$0.1477$ & $0.1480$ & $0.1478$ & $0.1497$ \\
 & EPOS-LHC & $0.1473$ & $[0.1465,0.1482]$ & $0.1469$--$0.1482$ & $0.1473$--$0.1484$ & $0.1480$ & $0.1475$ & $0.1493$ \\
 & SIBYLL-2.3d & $0.1488$ & $[0.1479,0.1497]$ & $0.1488$--$0.1494$ & $0.1488$--$0.1505$ & $0.1490$ & $0.1493$ & $0.1511$ \\
\midrule
$14$--$15$ & QGSJET-II-04 & $0.1482$ & $[0.1472,0.1494]$ & $0.1478$--$0.1484$ & $0.1478$--$0.1494$ & $0.1485$ & $0.1478$ & $0.1499$ \\
 & EPOS-LHC & $0.1480$ & $[0.1470,0.1490]$ & $0.1476$--$0.1484$ & $0.1480$--$0.1489$ & $0.1487$ & $0.1487$ & $0.1502$ \\
 & SIBYLL-2.3d & $0.1492$ & $[0.1481,0.1504]$ & $0.1488$--$0.1496$ & $0.1492$--$0.1496$ & $0.1491$ & $0.1494$ & $0.1511$ \\
\midrule
$15$--$16$ & QGSJET-II-04 & $0.0861$ & $[0.0854,0.0868]$ & $0.0861$--$0.0865$ & $0.0860$--$0.0876$ & $0.0862$ & $0.0860$ & $0.0898$ \\
 & EPOS-LHC & $0.0851$ & $[0.0843,0.0858]$ & $0.0847$--$0.0851$ & $0.0850$--$0.0865$ & $0.0848$ & $0.0851$ & $0.0885$ \\
 & SIBYLL-2.3d & $0.0866$ & $[0.0859,0.0874]$ & $0.0866$--$0.0867$ & $0.0866$--$0.0887$ & $0.0870$ & $0.0870$ & $0.0902$ \\
\midrule
$16$--$17$ & QGSJET-II-04 & $0.0603$ & $[0.0588,0.0616]$ & $0.0601$--$0.0608$ & $0.0603$--$0.0634$ & $0.0606$ & $0.0601$ & $0.0679$ \\
 & EPOS-LHC & $0.0586$ & $[0.0560,0.0611]$ & $0.0585$--$0.0606$ & $0.0586$--$0.0623$ & $0.0605$ & $0.0592$ & $0.0685$ \\
 & SIBYLL-2.3d & $0.0628$ & $[0.0608,0.0650]$ & $0.0622$--$0.0628$ & $0.0617$--$0.0661$ & $0.0632$ & $0.0620$ & $0.0713$ \\
\bottomrule
\end{tabular}
\end{minipage}\par\medskip

\par\medskip\noindent\begin{minipage}{\linewidth}
\centering\footnotesize\setlength{\tabcolsep}{2.6pt}
\captionof{table}{Accuracy of the floor estimate in the KASCADE geometry. For each sample the real nine
features are kept and the target is replaced by $g(X)+s(X)\,\xi$, with $g$ and $s$ learned from the
real target (gradient boosting for the conditional mean and for the squared residual) and $\xi$ standard
normal, drawn once per CORSIKA shower; the true floor $\sqrt{\langle s^2\rangle}$ is then known. Mean
bias of the estimate in per cent over 20 noise replicates for each setting of
table~\ref{tab:floorsens}, and the coverage of the reference setting's 95\% shower interval}
\label{tab:floorbench}
\begin{tabular}{llccccccccc}
\toprule
 & & & \multicolumn{7}{c}{bias of the floor estimate (\%)} & coverage \\
$\lg(E/\mathrm{eV})$ & model & true floor & reference & $K=5$--$50$ & robust & rank & Manhattan & six coord. & 1-NN & of interval \\
\midrule
$14$--$15$ (stations) & QGSJET-II-04 & $0.1481$ & $+0.3$ & $+0.3$ to $+0.5$ & $+0.2$ & $+0.6$ & $+0.5$ & $+0.3$ & $+1.6$ & $85\%$ \\
 & EPOS-LHC & $0.1479$ & $+0.3$ & $+0.3$ to $+0.5$ & $+0.3$ & $+0.6$ & $+0.4$ & $+0.2$ & $+1.6$ & $80\%$ \\
 & SIBYLL-2.3d & $0.1496$ & $+0.2$ & $+0.0$ to $+0.4$ & $+0.1$ & $+0.6$ & $+0.4$ & $+0.1$ & $+1.4$ & $90\%$ \\
\midrule
$14$--$15$ & QGSJET-II-04 & $0.1486$ & $+0.4$ & $+0.4$ to $+0.5$ & $+0.4$ & $+0.8$ & $+0.5$ & $+0.3$ & $+1.8$ & $85\%$ \\
 & EPOS-LHC & $0.1484$ & $+0.1$ & $-0.0$ to $+0.4$ & $+0.0$ & $+0.5$ & $+0.3$ & $+0.2$ & $+1.6$ & $95\%$ \\
 & SIBYLL-2.3d & $0.1495$ & $+0.4$ & $+0.2$ to $+0.4$ & $+0.3$ & $+0.8$ & $+0.5$ & $+0.4$ & $+1.6$ & $85\%$ \\
\midrule
$15$--$16$ & QGSJET-II-04 & $0.0870$ & $+0.5$ & $+0.2$ to $+0.5$ & $+0.4$ & $+2.0$ & $+0.6$ & $+0.4$ & $+4.4$ & $95\%$ \\
 & EPOS-LHC & $0.0849$ & $+0.6$ & $+0.6$ to $+0.6$ & $+0.6$ & $+2.1$ & $+0.6$ & $+0.3$ & $+4.9$ & $60\%$ \\
 & SIBYLL-2.3d & $0.0874$ & $+0.5$ & $+0.4$ to $+0.6$ & $+0.4$ & $+2.1$ & $+0.6$ & $+0.3$ & $+4.5$ & $90\%$ \\
\midrule
$16$--$17$ & QGSJET-II-04 & $0.0617$ & $+1.7$ & $+1.4$ to $+2.1$ & $+1.2$ & $+6.8$ & $+2.4$ & $+2.4$ & $+13.8$ & $75\%$ \\
 & EPOS-LHC & $0.0597$ & $+1.7$ & $+0.9$ to $+3.0$ & $+1.5$ & $+6.7$ & $+2.2$ & $+1.7$ & $+17.6$ & $90\%$ \\
 & SIBYLL-2.3d & $0.0627$ & $+2.4$ & $+2.4$ to $+4.1$ & $+1.9$ & $+7.5$ & $+2.4$ & $+2.5$ & $+17.1$ & $70\%$ \\
\bottomrule
\end{tabular}
\end{minipage}\par\medskip

\section{Measured muon-size distributions}
\label{app:measured}

\begin{figure}[H]
\centering
\includegraphics[width=0.859\textwidth]{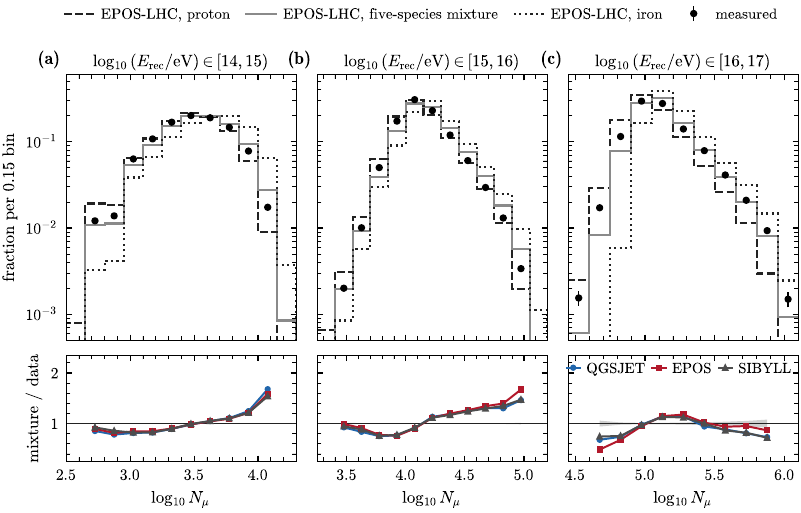}
\caption{In every decade of reconstructed energy the measured mean muon size lies between those of the proton and iron simulations, on all three models (table~\ref{tab:measured}). $\log_{10}N_\mu$ in three decades ((a)--(c)) of measured KASCADE events (black, Poisson errors) and of EPOS-LHC protons (dashed), the five-species mixture (grey) and iron (dotted), reweighted to a knee spectrum, after the common cuts. Lower panels of (a)--(c): mixture divided by data for all three models, where the data fraction exceeds $2\times10^{-3}$; grey band: Poisson error of the data; simulation errors not shown}
\label{fig:measuredhist}
\end{figure}

Figure~\ref{fig:measuredhist} compares the measured distribution of $\log_{10}N_\mu$ with the EPOS-LHC simulation in each decade of reconstructed energy; section~\ref{sec:measured} summarises it. Its
bins are $0.15$ wide because the distribution has fine structure at smaller widths.

\section{Station-image encoders}
\label{app:encoders}

The small CNN works on the $16\times16$ station grid with four channels (electromagnetic deposit,
muon deposit, arrival time, hit flag) and is trained for a fixed number of epochs chosen in advance.
The residual network and the station transformer take the same input. The graph network treats the
252 stations as nodes with those four signals and their position relative to the reconstructed core
($\Delta x$, $\Delta y$ and distance), passes messages four times between each station and its eight
nearest neighbours, and pools with attention. The three larger encoders are stopped on a validation set
of training showers. For table~\ref{tab:encoders} the graph network had $40$ epochs, the residual
network $30$ and the transformer $12$; a first graph-network run limited to $20$ epochs had still been
improving at its end, which is why it was given $40$, and in the $40$-epoch runs the best validation
epoch came well before the last (both training logs are kept in the analysis records). Used as a correction to gradient boosting on the nine features, each is
initialised to output zero.

Table~\ref{tab:encoders} is for QGSJET-II-04 with $\lgE\in[14,15]$, and gives means and ranges over
seeds (table~\ref{tab:stations} gives seed ensembles). In correction mode the network learns the
residual of gradient boosting on the nine features, whose predictions on the training showers are
computed out of fold. All networks use AdamW with weight decay $10^{-4}$, batch size $512$ and a cosine
learning-rate schedule (initial rate $10^{-3}$ for the CNN, $3\times10^{-4}$ for the residual network,
$5\times10^{-4}$ for the transformer and the graph network); the CNN is trained for $20$ epochs, and the
validation set for early stopping is a tenth of the training showers. The last row of
table~\ref{tab:encoders} is the same graph network given absolute station positions instead of
positions relative to the reconstructed core (section~\ref{sec:stations}).

\paragraph{The pre-registered repeats.} Table~\ref{tab:b2} and the larger run of
section~\ref{sec:stations} were fixed in \texttt{PREREG\_v3.md} before training and were run with one
version of the training script. Every encoder is trained for $20$ epochs, the three larger ones with
early stopping as above, with five seeds (three for the larger run, plus one seed of the shuffled
control). The $20$ epochs are the script's default, not the budgets of table~\ref{tab:encoders}; we
noticed this only after the runs. With $20$ epochs, $24$ of the $30$ graph-network corrections had
their best validation epoch at $19$ or $20$. On QGSJET-II-04 with $\lgE\in[14,15]$, where both budgets
were run on the same test events, the graph network's correction reaches $\sigma=0.1425$--$0.1427$ per
seed with $20$ epochs and $0.1425$--$0.1426$ with $40$, so the extra epochs did not help there. Shares are resolved to about $0.2$ points in $\lgE\in[14,15]$ and about $0.5$ in $[15,16]$, because $\sigma$ is stored to four decimals. Shares in table~\ref{tab:b2} are means of per-seed shares, with
$\sigma_{\rm nine}$ and $\sigma_{\rm oracle}$ from the same test events as table~\ref{tab:stations}.

\par\medskip\noindent\begin{minipage}{\linewidth}
\centering\footnotesize
\captionof{table}{Encoders of the simulated KASCADE station image. ``Image only'' is the encoder alone;
``correction'' is the encoder as a zero-initialised correction to gradient boosting on the nine
features ($0.1485$). Mean and range over seeds (three for the CNN of section~\ref{sec:kascade}, two
otherwise). The CNN is reported at a final epoch fixed in advance; the larger encoders are
stopped on a validation set of training showers. The last column is the share of the variance
gap to the MC-truth oracle that the correction closes.}
\label{tab:encoders}
\begin{tabular}{lcccccc}
\toprule
Encoder & size & image only & range & correction & range & gap closed \\
\midrule
small CNN & 0.2\,M & $0.1474$ & $0.1469$--$0.1480$ & $0.1432$ & $0.1431$--$0.1433$ & $10\%$ \\
residual network & 6.3\,M & $0.1458$ & $0.1458$--$0.1458$ & $0.1435$ & $0.1435$--$0.1435$ & $9\%$ \\
station transformer & 1.2\,M & $0.1459$ & $0.1459$--$0.1459$ & $0.1431$ & $0.1431$--$0.1431$ & $10\%$ \\
graph network & 0.4\,M & $0.1432$ & $0.1432$--$0.1433$ & $0.1426$ & $0.1425$--$0.1426$ & $11\%$ \\
graph net., absolute pos. & 0.4\,M & $0.1431$ & $0.1431$--$0.1432$ & $0.1424$ & $0.1424$--$0.1425$ & $11\%$ \\
\bottomrule
\end{tabular}
\end{minipage}\par\medskip

\FloatBarrier
\section{Shower identity}
\label{app:showerkey}

The analysis identifies a shower by its true energy and primary. Stored in single precision, that key is not unique: in the QGSJET-II-04 KASCADE set, $\keyOneMergedPct\%$ of the showers between $10^{14}$ and $10^{15}$\,eV share it with at least one other shower, and $\keyTwoMergedPct\%$ and $\keyThreeMergedPct\%$ in the next two decades (about a third in the lowest decade of EPOS-LHC and SIBYLL-2.3d, and at most $\keyCombMax\%$ in any KASCADE-Grande set). Such showers stay together in every split and are never used as each other's neighbours. That can only reduce leakage, but it slightly raises the floor estimate. We reran the QGSJET-II-04 KASCADE decades $10^{14}$--$10^{16}$\,eV and the QGSJET-II-04 KASCADE-Grande set with showers keyed also by their true particle numbers; these reruns also redraw the split. The floor estimate falls by at most about $\keyFloorMax\%$ and boosting's excess moves by at most $\keyExMax$ points. Between $10^{15}$ and $10^{16}$\,eV the excess becomes $\keyTwoEx\%$, with an interval ($\keyTwoExLo$ to $\keyTwoExHi\%$) that excludes zero, still within the $2.2\%$ of section~\ref{sec:plateau}; no pre-registered criterion or stated conclusion changes.
\end{appendices}

\renewcommand{\bibfont}{\small}
\bibliography{epjc}

\end{document}